# Ultra-high THz-field-confinement at $LaAlO_3$ twin walls

J. Wetzel[1,5*], J. Taboada-Gutiérrez[2*a], M. Roeper[1], F.G. Kaps[1], G. Esposito[2], D. Marchese[2], R. Buschbeck[1], P. Lenz[1], J.M. Klopf[3], H. A. Bechtel[4], S. N. Gilbert Corder[4], J. Teyssier[2], S.C. Kehr[1], L.M. Eng[1,5], A. B. Kuzmenko[2], and S.D. Seddon[1a]

[1]Institute of Applied Physics, TUD Dresden University of Technology, Nöthnitzer Straße 61, 01187 Dresden, Germany
[2]University of Geneva, Department of Quantum Matter Physics (DQMP), Geneva 1211, Switzerland
[3]Institute of Radiation Physics, Helmholtz-Zentrum Dresden Rossendorf, 01328 Dresden, Germany
[4]Advanced Light Source (ALS), Lawrence Berkeley National Laboratory, Berkeley, CA 94720, USA
[5]Würzburg-Dresden Cluster of Excellence (EXC2147) ctd.qmat – Complexity, Topology, and Dynamics in Quantum Matter, Dresden, Germany

*These authors contributed equally
[a]email: samuel.seddon@tu-dresden.de, and javier.taboadagutierrez@unige.ch

## Abstract

The control and steering of light at nanometre length scales is crucial for the development of both fundamental science and nanophotonic technologies. Recent advancements have been achieved by exploiting various crystalline anisotropies, allowing for subdiffractional and diffraction-less canalisation of energy. These studies in particular benefit from stacking and twisting of 2D materials, whereas corresponding capabilities of anisotropic bulk crystals are rather unexplored. In this work, we show that ferroelastic twin walls - crystallographically perfect 2D-sheets that separate regions of differently oriented domains – in the distorted perovskite $LaAlO_3$ provide a natural platform for broadband lateral confinement and superb canalisation of light at the nanoscale. Without fabrication processes, the electromagnetic fields localised at such walls exhibit lateral optical sizes up to 260 times smaller than the free-space wavelength. Depending on the adjacent domain orientation and frequency, the twin wall pattern preferentially concentrates or repels the electromagnetic energy, constituting a natural building block towards broadband MIR and THz nanophotonics for polaritonic circuitry.

## Introduction

The control, manipulation and steering of light at nanometre length scales is a central goal of nano-optics [1, 2, 3, 4]. Anisotropic polar crystals of vdW materials [5, 6, 7] provide an attractive platform because they support phonon polaritons (PhPs) [8], hybrid light-matter modes formed by coupling electromagnetic waves to lattice vibrations in polar insulators. In hyperbolic regimes [9]—where different components of the dielectric permittivity tensor have opposite signs—PhPs can propagate highly directional with deeply sub-wavelength confinement [10, 11]. A particularly promising application is PhP canalisation [12, 13, 14, 15, 16, 17, 18, 19], in which energy flow is concentrated along specific directions with diffraction-less propagation.

In most current approaches, canalisation is engineered through the stacking and twisting of layers of polar materials [2 , 16, 17, 18, 19] and then adding nanopatterned elements (e.g. e-beam/FIB-defined launchers, reflectors, or resonators) to control PhP excitation. Natural canalisation as reported in pristine crystals [20] is typically restricted to narrow spectral windows close to phonon resonances, suffering from limited propagation lengths and diffraction. Consequently, a platform that provides strong confinement and long-range propagation while avoiding complex fabrication addresses key limitations of existing state-of-the-art nano-optical devices and will be highly relevant for THz/MIR nano-optics and spectroscopy.

In this work, we introduce ferroelastic twin walls (TWs) as a suitable and effective platform for canalizing light at the nanoscale without the necessity of complex fabrication processes. These so-called ferroelastic materials undergo a spontaneous crystal symmetry breaking below a certain ordering temperature (on the order of hundreds of ℃), resulting in several possible crystallographic domains, that co-exist and that tessellate throughout the material [21], meeting at mirror planes. These regions are delineated by TWs, which form a well-defined pattern in the material and constitute non-polar, uncharged interfaces aligned along specific crystallographic angles, separating two domains of the same material with distinct orientations of their anisotropy axis. Furthermore, as TWs form in certain symmetry-defined planes and represent a perfect boundary between two ferroelastic domains, they therefore can be thought of as atomically sharp, pristine interfaces. Ferroelastic twinning is a common phenomenon in nature, which is notably present in various minerals and materials such as calcite ($CaCO_3$), quartz ($SiO_2$), feldspar, or lanthanum aluminate ($LaAlO_3$, LAO), as used here. Indeed feldspars alone make up 30% of the earth's crust [22].

LAO is one of the most versatile complex oxides for functional interfaces and heterostructures, as it can act both as a polar active layer and as epitaxial substrate. In the former case, it has been successfully employed for studying and controlling high mobility 2D electron gases [23, 24, 25], and for the control of quantum phases and quantum circuitry [26, 27]. In the latter case, LAO stabilizes strained phases of oxide films such as $BiFeO_3$ [28, 29] or $NdNiO_3$ [30], for enhancing their multiferroic behaviour or controlling edge-confined polaritons in the epitaxially grown thin film layer, respectively, among other applications. Taken together, these properties make LAO a powerful platform for engineering functional devices, nanoscale control of ferroic order and enhanced light-matter coupling in thin films.

Using scattering scanning near-field optical microscopy (s-SNOM), analytical dipole modelling, and finite-element method (FEM) simulations, we demonstrate here ultra-high electromagnetic lateral field confinement at LAO TWs in the technologically relevant terahertz (THz) and mid-infrared (MIR) ranges. As a function of the excitation frequency, we demonstrate that TWs form a tuneable network across the crystal for the canalisation of light at the nanoscale, with frequency-selective paths for hosting the enhanced near field.

## Results

**Optical anisotropy and twin-wall pattern.** In this study LAO, a rhombohedrally distorted pseudocubic perovskite (space group $R\bar{3}c$) is investigated [31]. LAO is optically uniaxial, so its permittivity can be decomposed into two components, $\varepsilon_{\parallel}$ and $\varepsilon_{\perp}$, defined parallel and perpendicular to the optical axis aligned with the pseudocubic $[111]$ direction (coincident with the spontaneous compressive strain axis). In the MIR–THz range the dielectric function exhibits three high-reflectivity Reststrahlen bands (RBs) [32, 33, 34], where $\mathrm{Re}(\varepsilon)$ becomes negative and PhPs can be excited. Figure 1a shows the real parts of the uniaxial dielectric permittivity components of LAO, $\varepsilon_{\parallel}$ and $\varepsilon_{\perp}$, extracted from polarised Fourier transform infrared spectroscopy (FTIR) (see Supplementary Information S1). Notably, the three RBs for each of the permittivity components, $\varepsilon_{\parallel}$ and $\varepsilon_{\perp}$, show a small frequency shift between their corresponding phonon resonances, introducing an elliptic or hyperbolic response within all RBs.

As all ferroelastic crystals, LAO forms multiple symmetry-related crystallographic domains. On LAO(001), four domain variants are allowed and can be visualised by taking a cube distorted along its $[111]$ direction (pink domain labelled “I” in Fig. 1b) and rotating it by $90°$ in-plane three successive times to obtain domains II–IV. While some domain pairs share the same in-plane projection of the optical axis, their out-of-plane projections align mirror-symmetric with respect to the $[001]$ direction, which becomes important when interpreting near-field contrast, as will be shown later.

By crystal symmetry, the four domain variants are able to form six different TW orientations in bulk rhombohedral LAO, but only four occur at the $(001)$ surface (Fig. 1c) [21, 31]: long vertical $(010)$ TWs, intersected by the zig-zag $(110)-(1\bar{1}0)$ TW set (which operate at $45°$ from the pseudocubic directions). This arrangement produces the characteristic “chevron” tiling commonly observed in ferroelastic crystals. Note that the domains shown in Fig. 1b do not tessellate in that rotational order. Instead, in order to ensure TW formation which is strain compatible, in accordance with Sapriel [35], domains alternatingly tessellate vertically in pairs of I-III and II-IV, as seen in Fig. 1c. A (later critical) consequence of this is, that whilst moving along vertical pairs (and therefore across diagonal TWs), the optical axes are oriented along the same in-plane direction, with however opposing out-of-plane components. Additionally, the optical axes along a set of 45°-walls rotate in plane, with the out-of-plane component staying unchanged. In Fig. 1c the projections of the optical axes onto the $(001)$ surface is indicated as grey arrows, with arrowheads pointing out of the plane. This leads to effective head-to-head and tail-to-tail walls. All TWs in this orientation run perfectly perpendicular to the surface plane and extend through the full sample thickness. TWs meet always at twin-wall junctions (TWJs) where all four domains and four walls intersect (see dashed boxes in Fig. 1c).

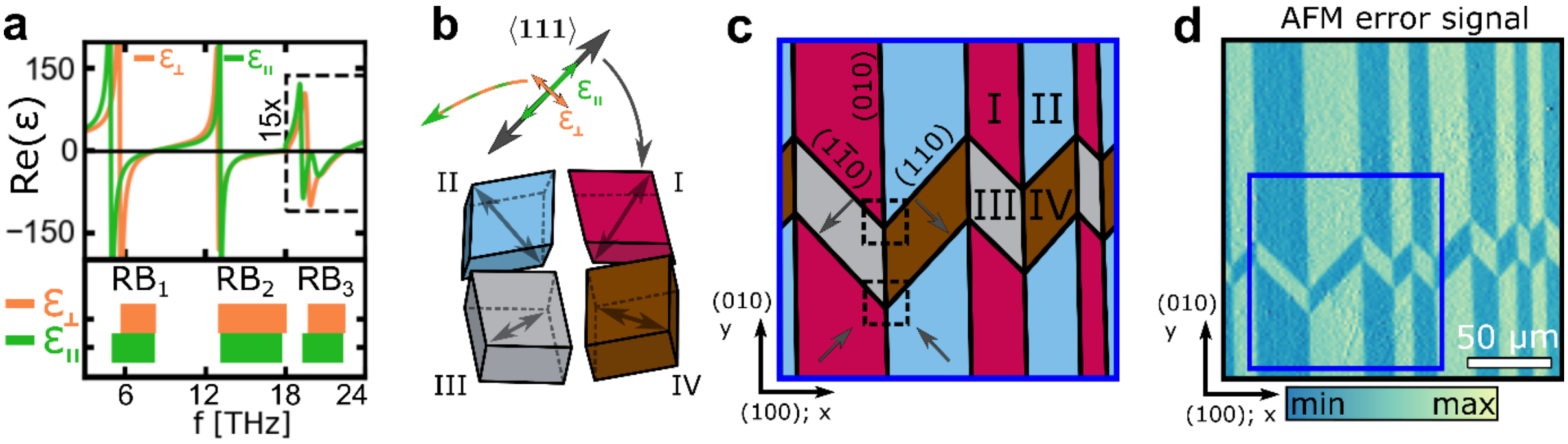

**Figure 1 | $LaAlO_3$ properties and twin domains. a** (upper) Anisotropy in the dielectric permittivity $\varepsilon$ decomposed into perpendicular (orange, $\varepsilon_{\perp}$) and parallel (green, $\varepsilon_{\parallel}$) components as defined with respect to the LAO optical axis. (lower) The three Reststrahlen bands spectral positions shown for the same permittivity components. **b** (top) Schematic indicating the orientations of $\varepsilon_{\parallel}$ and $\varepsilon_{\perp}$ relative to the strain axis (grey arrow), coinciding with the pseudocubic $[111]$ direction. (bottom) Sketch of the four pseudocubic crystallographic domains of LAO (I-IV), compressively distorted along the $[111]$ direction. Domains II–IV are $90°$ rotations of domain I. **c** The allowed tiling of a LAO(001) surface following the crystal twinning rules. The optical axes of the four domains are indicated by grey arrows pointing out of plane. Vertical and $45°$ TWs are labelled according to their Miller indices and their possible TWJs are marked with dashed boxes. **d** Atomic force microscopy (AFM) error signal image indicating relative tilts of the sample surface and the twin domains. The blue box indicates the sketched area in panel **c**.

The domain arrangement is directly observed by atomic force microscopy. The AFM error-signal image shown in Fig. 1d (effectively the lateral derivative of the height topography) highlights regions of constant surface tilt of about ±0.075° [21] associated with the rhombohedral angles of the four domains, yielding a terraced surface morphology consistent with the schematic in Fig. 1c.

**THz nano-imaging of twin walls.** To determine the near-field response at TWs, THz-s-SNOM measurements have been performed utilising a tuneable free-electron laser (FEL) as narrowband light source [36] (FEL-SNOM, see Fig. 2a and Methods). Figure 2b shows a FEL-SNOM image $s_2(x,y)$ (second harmonic demodulation of the tip tapping frequency) at $f_2 = 8.07\ \mathrm{THz}$ ($\lambda_0 \approx 37.15\ \mu\mathrm{m}$, $\omega_0 \approx 269.2\ \mathrm{cm}^{-1}$) in a sample region containing several TWs and TWJs. This frequency lies within $RB_1$ close to the LO phonon, where both $\varepsilon_{\parallel}$ and $\varepsilon_{\perp}$ are slightly negative and highly dispersive (Fig. 2c). The near-field map exhibits a set of bright and dark contrasts (relative to the bulk LAO signal) localised at TWs positions.

A striking feature is that the near-field contrast at the TWs depends on the orientation of the neighbouring LAO domains. Any vertical $(010)$ TW experiences a contrast inversion when crossing a TWJ, whereas the zig-zag TWs remain consistently bright or consistently dark along their length, depending in all cases on whether the optical axis orientation in the adjacent walls points toward or away from the TW. This repeatable and robust pattern defines two easily identifiable junction types, $TWJ_1$ and $TWJ_2$ (blue and black rectangles in Fig. 2b); two further junctions ($TWJ_1'$, $TWJ_2'$) are $180°$-rotated counterparts and behave equivalently.

**Frequency-dependent contrast inversion and experimental confinement.** To probe the spectral response and lateral field confinement, $TWJ_1$ (blue square in Fig. 2b) was imaged at two nearby frequencies, $f_1 = 7.96\ \mathrm{THz}$ ($\lambda_0 \approx 37.66\ \mu\mathrm{m}$, $\omega_0 \approx 265.5\ \mathrm{cm}^{-1}$) and $f_2 = 8.07\ \mathrm{THz}$ (vertical dashed lines in Fig. 2c). Strikingly, the bright–dark pattern of the junction inverts completely between these frequencies (Fig. 2d,e). Although only one RB is displayed here, a similar behaviour can be observed in all 3 RBs and can be found in the Supplementary Information S3. A line profile extracted from a high-resolution scan of $TWJ_1$ at $f_2$ (see inset in Fig. 2e), yields an optical $\mathrm{FWHM} = 143\ \mathrm{nm}$, corresponding to a lateral confinement factor of $260$ relative to the excitation wavelength (see Supplementary Information S3). However, considering that the measured $\mathrm{FWHM}$ is a convolution of the field confinement and tip radius, the true confinement factor is likely to be even higher as confirmed below by numerical simulations.

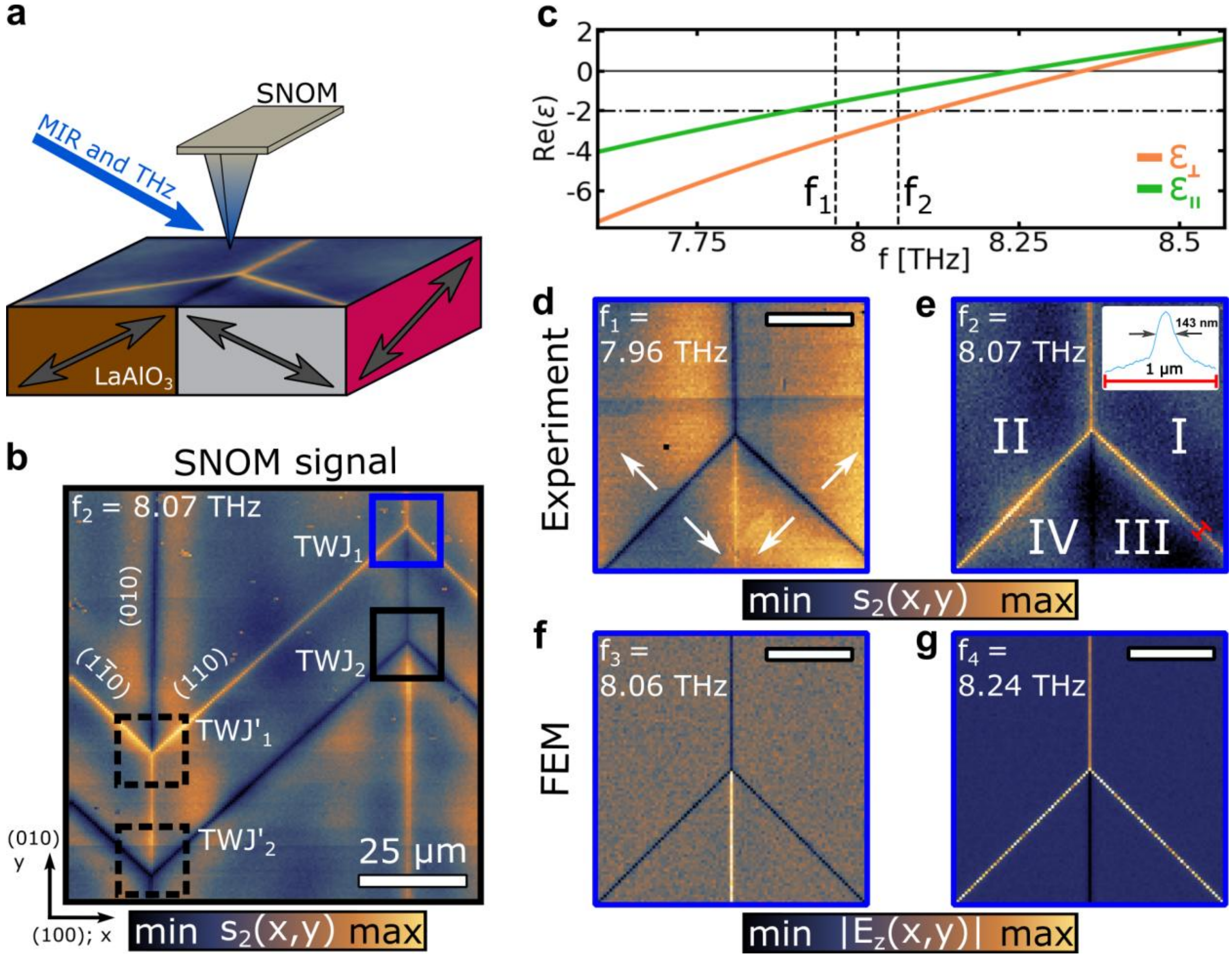

**Figure 2 | Frequency-dependent response of twin walls. a** Free-electron-laser-illuminated s-SNOM measurements reveal sharp optical contrast emergent from TWs, with the optical axis projections of the twin domains indicated as grey arrows. Incoming light ranges from the MIR to THz spectral regime. **b** SNOM image of all four possible TW junctions (TWJs), taken at an illumination frequency of $f_2 = 8.07\ \mathrm{THz}$. **c** Real part of the dielectric permittivity $\varepsilon$ in the relevant spectral regime, decomposed into perpendicular (orange, $\varepsilon_\perp$) and parallel (green, $\varepsilon_\parallel$) components, defined relative to the LAO optical axes. **d**, **e** s-SNOM images acquired at $TWJ_1$ as a function of illumination frequencies $f_1$ and $f_2$ (as specified in **d**, **e** respectively and marked in **c**). Inset in **e** shows the line profile across a bright (110) TW. **f**, **g** FEM simulations of the same $TWJ_1$ at frequencies $f_3$ and $f_4$, which are spectrally slightly shifted to match the contrast observed in the experiments (d, e). Scale bars in **d**-**g** correspond to $10\ \mu\mathrm{m}$.

**FEM simulations and origin of the contrast.** To understand the contrast inversion and the highly confined fields, the tip–sample interaction was calculated through FEM simulations, as it has successfully been employed in previous works to explain the image contrast in s-SNOM experiments [37, 38, 39, 40] (see Methods). In these simulations, a vertically-oriented point dipole is scanned laterally above the sample at $f_3 \approx 8.06\ \mathrm{THz}$ ($\lambda_0 \approx 37.17\ \mu\mathrm{m}$, $\omega_0 = 269.0\ \mathrm{cm}^{-1}$) and $f_4 \approx 8.24\ \mathrm{THz}$ ($\lambda_0 \approx 36.36\ \mu\mathrm{m}$, $\omega_0 = 275.0\ \mathrm{cm}^{-1}$) reproducing perfectly the bright/dark contrast of the measured $s_2(x, y)$ images at $f_1$ and $f_2$ (Fig. 2f,g versus Fig. 2d,e), including the strong field localisation at the walls. Note that the simulation surface is perfectly flat, demonstrating that the contrast is driven by the local in-plane rotation of the optical axis between the neighbouring domains and not by any topography-related effect. In fact, all s-SNOM measured samples have been polished for removing any such surface tilt. The spectral shift towards slightly higher frequencies is consistent with the frequency shift previously reported for s-SNOM simulations utilising the point-dipole approximation [41]. Intriguingly, the simulations even capture the small but systematic brightness difference between the vertical $(010)$ and the $45°$ zig-zag walls (Fig. 2e,g), which follows from neighbouring optical axes meeting at an in-plane angle of $90°$ (for $(010)$ walls) or $180°$ (for the $45°$ zig-zag walls). Merely the weak experimental intensity gradients next to some walls are not reproduced. This is attributed to the convolution of 2 effects: surface-PhPs launched by the tip and weakly reflected by the TW, and illumination-related effects. In fact, in complementary synchrotron infrared nano-spectroscopy (SINS) measurements the excitation of non-strongly confined PhPs at the surface of bulk LAO(001) has been confirmed in this spectral regime (see Methods and Supplementary Information S2).

**Modified dipole model and optical contrast mechanism at the walls.** To further elucidate why different walls brighten or darken at different frequencies, the analytical point-dipole model commonly used to describe s-SNOM contrast was adapted [36, 37, 42, 43]. Specifically, extending it to treat a vertical interface (orthogonal to the surface) separating two adjacent LAO domains described by fully general permittivity tensors with arbitrarily oriented optical axes (for details see SI S4). While an analytical treatment of an entire TWJ is highly non-trivial, modelling a single TW is found to be sufficient to explain the contrast of all TWs present in a single TWJ. We therefore apply the model to all TW types present on LAO(001) (Fig. 3a) and introduce the notation $\mathrm{TW}_{v/d}^{\parallel/\perp}$, where $v$ (vertical) and $d$ (diagonal) denotes the wall direction. Moreover, the superscript ∥ (⊥) distinguishes the set of walls that appear bright at $f_1$ ($f_2$) in the experiments as summarised in Fig. 3b, i.e. when the parallel (perpendicular) dielectric component, $\varepsilon_\parallel(\varepsilon_\perp)$, approaches the Mie resonance condition $\mathrm{Re}(\varepsilon_\parallel) \approx -2$ ($\mathrm{Re}(\varepsilon_\perp) \approx -2$) [44] (dashed-dotted horizontal line in Fig. 2c and Supplementary Information S4). Since the $(110)$ and $(1\bar{1}0)$ walls within one zig-zag set behave equivalently, we group them as $\mathrm{TW}_d$. The grey arrows in Fig. 3a (optical-axis projections) then motivate two geometrical classes: $\mathrm{TW}^\parallel$ (i.e. $\mathrm{TW}_v^\parallel$ and $\mathrm{TW}_d^\parallel$), where the optical axis points toward the TW–surface junction, and $\mathrm{TW}^\perp$ (i.e. $\mathrm{TW}_v^\perp$ and $\mathrm{TW}_d^\perp$), where it points toward the wall but away from the surface. For example, $\mathrm{TWJ}_1$ consists of $\mathrm{TW}_v^\perp$, two $\mathrm{TW}_d^\perp$, and $\mathrm{TW}_v^\parallel$ (the remaining junction compositions are listed in Supplementary Information S3), and Fig. 3b,c summarizes which TW branches are bright at $f_1$ and $f_2$, respectively.

In the adapted analytical model, the near-field probe is represented by a small metallic sphere with isotropic polarizability $\alpha_t$, located at height $h$ above the surface (exemplarily shown in Fig. 3c for $\mathrm{TW}_d^\perp$ separating e.g. domains III and I). Moreover, each adjacent LAO domain is represented by an induced dipole, that is determined by the local anisotropic permittivity tensor and optical-axis orientation via its anisotropic polarizability tensor (e.g. $\hat{\alpha}_{\mathrm{I}}$, $\hat{\alpha}_{\mathrm{III}}$). For s-SNOM at the TW, it is the tip field that induces these dipoles in the two domains at the TW–surface junction, resulting in a TW-type dependent near-field coupling (see Supplementary Information S4 for details). Note that the polarizability tensors in the examined strongly dispersive regime of LAO are highly frequency dependent as indicated by the differently oriented ellipsoids in Fig. 3c. The resulting calculated scattering cross sections for all four TW types within $\mathrm{RB}_1$ consequently shows a highly frequency-dependent response (Fig. 3d), perfectly reproducing the experimentally and numerically observed contrast inversion: $\mathrm{TW}^\parallel$ exhibits a maximum near $f_1$ while $\mathrm{TW}^\perp$ is $\approx 25\%$ smaller, whereas near $f_2$ the response reverses and $\mathrm{TW}^\perp$ becomes maximal while $\mathrm{TW}_\parallel$ is $\approx 55\%$ smaller. The spectral positions of these maxima in the model coincide with resonances of the tensorial domain polarizability components ($\alpha_\parallel$, $\alpha_\perp$), resulting in a simple rule for observing an enhanced TW response: a wall appears bright when the dielectric component that points toward the TW-surface junction, i.e. $\mathrm{Re}(\varepsilon_\parallel)$ or $\mathrm{Re}(\varepsilon_\perp)$, approaches $-2$ (Fig. 2c and Supplementary Information S4). Note that even though this simple analytical approximation is suitable to explain the spectral response of the TWs, for a full theoretical description of the near-field images observed, including the TWJ and the adjacent domains, numerical simulations are required.

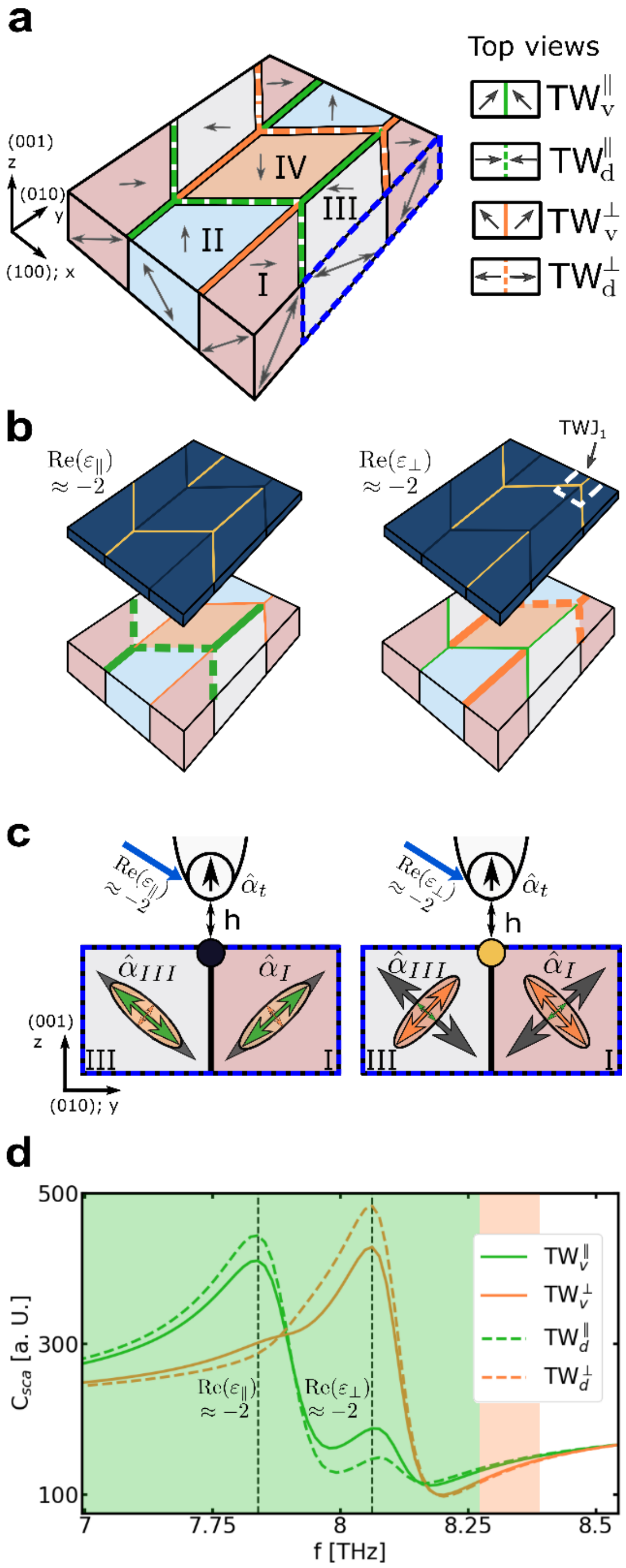

**Figure 3 | Contrast mechanism at the twin walls. a** Schematic showing the four types of TWs present in the system, characterised by two groups marked orange and green. Orange (green) TWs show an enhanced response when the perpendicular (parallel) permittivity component is resonantly excited. Optical axis projections are marked as double arrows on the sides, while single arrows indicating their surface projection direction with an out-of-plane component (arrowhead pointing out of the surface, compare Fig. 1c). **b** lower row: TWs selected for enhanced fields at frequencies where $\mathrm{Re}(\varepsilon_{\parallel})$ and $\mathrm{Re}(\varepsilon_{\perp})$ approach $-2$, respectively, and the subsequent TW network pattern above each (upper row). **c** Schematic indicating relevant parameters for the adapted dipole model assuming dipole excitation in the tip ($\alpha_t$) as well as in the domains at either side of the TW ($\hat{\alpha}_I$, $\hat{\alpha}_{III}$ in the example shown) according to their frequency-dependent permittivity tensor orientation (ellipsoids in I, III). **d** Calculated scattering signatures for the four types of TWs with maxima occurring for $\mathrm{Re}(\varepsilon_{\parallel/\perp}) \approx -2$ (compare Fig. 2c) and contrast inversion apparent at 7.88 THz.

**Depth-resolved confinement and energy channelling.** Finally, FEM using a fixed dipole position was applied to analyse how the near-field distribution evolves above and below the surface and to quantify the ultimate field confinement. At $f_3$ (Fig. 4a), the near-field distribution in the surface plane (XY plane) exhibits multi-lobe features within each domain due to the tilted and rotated optical axes, in clear contrast with an isotropic medium without domains (Fig. 4a inset), where no lobe features exist and the energy is isotropically radiated. Along $\mathrm{TW}_d^{\perp}$ the field concentrates near the junction but decays rapidly with distance, with $|E_z|$ values lower than those in the surrounding domains. Along $\mathrm{TW}_v^{\parallel}$ the field remains enhanced over larger distances, whereas $\mathrm{TW}_v^{\perp}$ shows suppression near the junction with only weak enhancement further away. Cross-sections in the XZ plane (Fig. 4b, indicated by the dashed lines termed Plane 1, for the upper panel, and Plane 2, for the lower panel, in Fig. 4a, taken $10\ \mu\mathrm{m}$ away from the dipolar source) show that above the surface the field is broadly distributed with a localised hot spot at the TW position, while below the surface a similar pattern appears with much lower field penetration. Corresponding line profiles $1\ \mathrm{nm}$ above the surface (Fig. 4c, indicated by the yellow arrow in Fig. 4b upper panel) quantify the TW-bounded electric field confinement, manifested as peaks/dips at the TW positions superimposed on an otherwise broad background attributed to a surface PhP mode on the domains.

At $f_4$ (Fig. 4d), the behaviour changes qualitatively: the near field accumulates strongly along the $TW^{\perp}$ branches and remains highly localised along the walls for many tens of microns away from the excitation point, with negligible lateral spreading. XZ cross-sections (Fig. 4e, at Plane 1 and Plane 2, upper and lower panels, respectively) reveal intense and highly localised hot spots at the surface intersection of $\mathrm{TW}_v^{\perp}$ (upper panel) and at $\mathrm{TW}_d^{\perp}$, while $\mathrm{TW}_v^{\parallel}$ is suppressed (lower panel). Line profiles $1\ \mathrm{nm}$ above the surface (Fig. 4f) yield a FWHM of $34.4\ \mathrm{nm}$ (for the Plane 1 case, upper panel) at the intersection between $\mathrm{TW}_v^{\perp}$ and the bulk LAO surface, corresponding to an unprecedented lateral confinement of $\lambda_0/1057$ for $\lambda_0 = 36.36\ \mu\mathrm{m}$, and similarly narrow maxima at $\mathrm{TW}_d^{\perp}$ (Plane 2, lower panel). Note that, this ultimate limit is inaccessible in the s-SNOM experiment, due to the limitation imposed by the finite size of the probe.

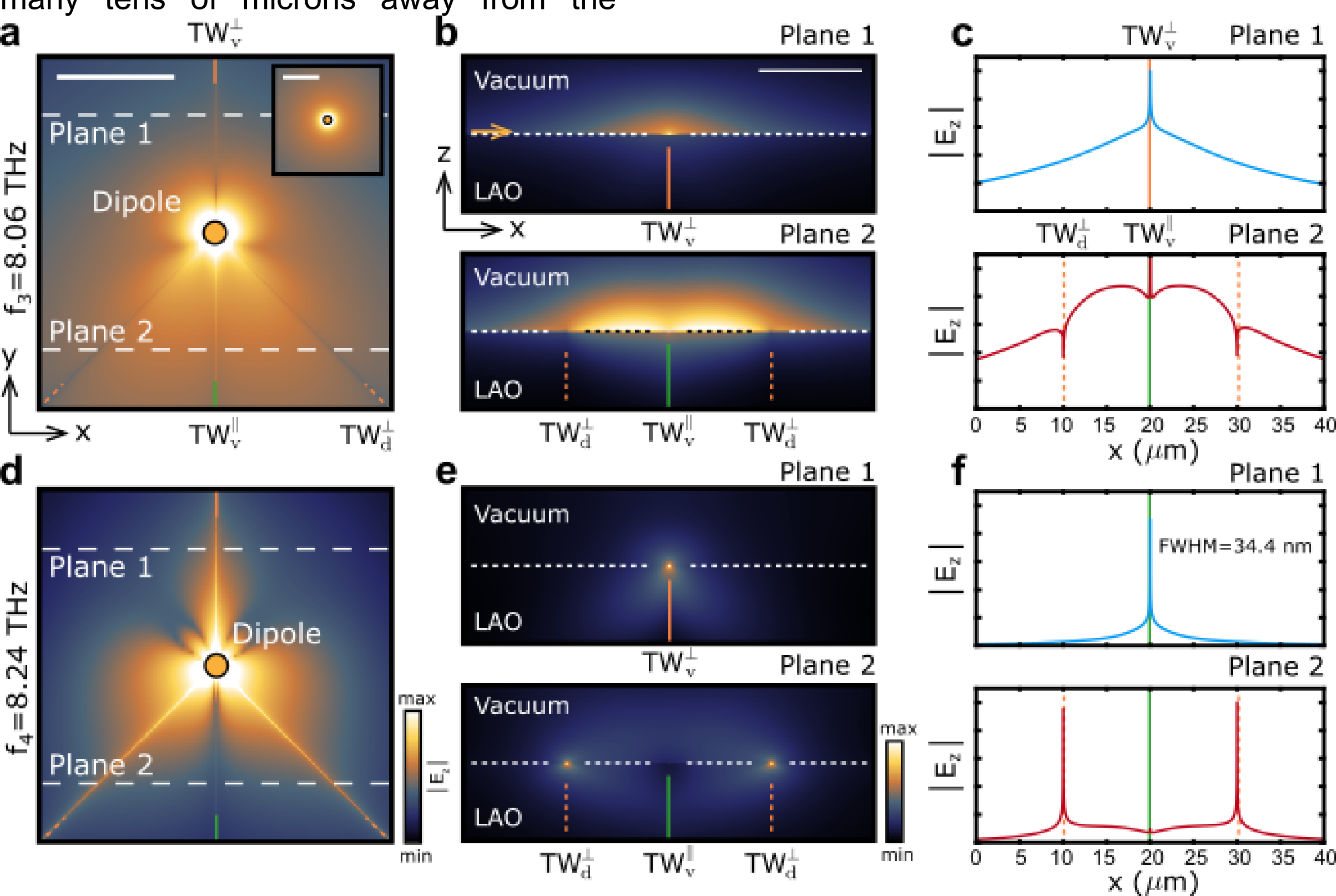

**Figure 4 | Numerical analysis of the optical confinement. a**, **d** Electric field distribution at the interfaces of LAO TWs at **a** $f_3 \approx 8.06\ \mathrm{THz}$ ($\lambda_0 \approx 37.17\ \mu\mathrm{m}$, $\omega_o = 269.0\ \mathrm{cm}^{-1}$), and **d** $f_4 \approx 8.24\ \mathrm{THz}$ ($\lambda_0 \approx 36.36\ \mu\mathrm{m}$, $\omega_o = 275.0\ \mathrm{cm}^{-1}$). The images represent a XY cut of the electric field distribution, $|E_z|$, in the XY plane, at $1\ \mathrm{nm}$ above the sample's surface. Scale bar represents $10\ \mu\mathrm{m}$. **b**, **e** Electric field distributions at $f_3$ (**b**) and $f_4$ (**e**) for XZ planes. Top panels represent a XZ cut taken along the top white dashed line in **a** and **d** (denoted as Plane 1) and bottom panels represent a XZ cut taken along the bottom white dashed line (denoted as Plane 2). Scale bar represents $10\ \mu\mathrm{m}$. **c**, **f** Line profiles of the field distribution taken $1\ \mathrm{nm}$ above the sample's surface, as denoted by the yellow arrow in **b** top panel at $f_3$ (**c**) and $f_4$ (**f**). Top panels represent profiles taken along Plane 1 and bottom panels represent profiles taken along Plane 2. A clear effect of the twin boundary is observed on the electric field distribution yielding a theoretical field confinement of up to $\lambda_0/1057$.

## Conclusion

In summary, we demonstrate experimentally and theoretically that ferroelastic twin walls in $LaAlO_3$ provide a pattern-free and intrinsically reconfigurable platform for concentrating and routing THz–MIR near fields at the nanometre length scale. The broken translational symmetry at a given TW, combined with the intrinsic optical anisotropy of LAO enables selected walls to act as channels of the near field. By tuning the illumination frequency, the TW contrast switches in a reproducible manner, effectively turning specific branches of the naturally occurring TW network "on" or "off" and directing highly confined near-field signals along predefined crystallographic directions.

Quantitatively, an experimental lateral confinement factor of $260$ ($\mathrm{FWHM} \approx 143\,\mathrm{nm}$ at $\lambda_0 = 37.17\,\mu\mathrm{m}$) was observed, while finite-element modelling predicts confinement down to $34.4\,\mathrm{nm}$, corresponding to $\lambda_0/1057$ at $\lambda_0 = 36.36\,\mu\mathrm{m}$, at selected TWs. Crucially, the TW-bound near field remains highly localised along the walls over tens of micrometres—i.e. over mesoscopic distances that exceed the lateral confinement by more than three orders of magnitude—while exhibiting negligible lateral spreading. This combination bridges long-range, guided near-field transport with deep-subwavelength lateral modal confinement in an unpatterned bulk dielectric.

More broadly, ferroelastic twinning is widespread and offers multiple external control knobs (e.g. mechanical strain, temperature; and, in ferroelectrics, electrical control), suggesting routes toward dynamically reconfigurable THz/MIR near-field routing. In addition, integrating ferroelastic TW networks with low-loss hyperbolic polaritonic materials (e.g. hBN, $\alpha$-$MoO_3$ or $\alpha$-$V_2O_5$) could extend the concept toward compact polaritonic circuitry and enhanced nanospectroscopy/nanochemistry based on strongly confined, directional near fields.

## Methods

*Sample description*

The samples studied within this work are (001)-grown bulk LAO crystals by Crystal GmbH, Germany. On LAO(001), four domains are naturally formed with a rhombohedral angle between sub-facets of only $\approx 0.075°$, so the surface is effectively flat. Additionally, the overall crystallographic angles can be suppressed by a further polishing process, which was performed on all samples measured in this study. The exact sample area investigated within this work was recharacterized AFM and Raman-measurements, enabling the determination if the exact axis orientation and domain identification.

*Fourier-Transform Infrared Spectroscopy*.

Polarised Fourier-transform infrared spectroscopy (FTIR) was employed to characterise the MIR-THz optical response of a (110)-grown bulk LAO substrate. FTIR spectra were collected by means of a bolometer detector (frequency range from $120$ to $600\,\mathrm{cm}^{-1}$) and a MCT detector (frequency range from $500$ to $7000\,\mathrm{cm}^{-1}$), employing a globar as a broadband light source. A layer of $100\,\mathrm{nm}$ of Au was deposited on part of the LAO substrate and used as reference for normalising the near-field signal. The angle of incidence of the incoming light on the sample was approximately $15°$. For details on the permittivity extraction see Supplementary Information S1.

*Scattering-Type Scanning Near-Field Optical Microscopy.*

Scattering-type scanning near-field optical microscopy (s-SNOM) measurements were performed by means of commercially available neaSCOPE s-SNOM systems (Attocube Systems GmbH). The systems employed either the radiation from a synchrotron (SINS) [45] or from a free-electron laser [36] as excitation light sources in the spectral regimes of 40-2000cm$^{-1}$ or $5 - 250\,\mu\mathrm{m}$, respectively.

The SINS measurements were carried out at the Advanced Light Source, ALS, at Lawrence Berkeley National Laboratory on a (001)-grown LAO crystal and were used to demonstrate surface PhP launching and to extract the dispersion (see Supplementary Information S2). SINS is based on the s-SNOM technique and provides nanoscale-resolved absorption spectra with wavelength-independent spatial resolution. In this case, we have employed a lithograpically-defined Au bar as polaritonic launcher. We acquired third-harmonic SINS line scans perpendicular to the Au bar (see Supplementary Information Fig. 9), which reveal enhanced near-field signal and sets of interference fringes within RB1 and RB2 (RB3 fringes are not resolved due to higher intrinsic losses), with frequency-dependent fringe period. We attribute this oscillations to the excitation of non-strongly confined surface PhPs at the surface of bulk LAO(001) (see Supplementary Information S2 for a full discussion).

FEL-SNOM has been carried out using the pulsed free-electron laser FELBE at the Helmholtz-Zentrum Dresden Rossendorf (HZDR), covering the full frequency range from 1.2 to 60 THz with 13 MHz repetition rate, a pulse length between 1 and 25 ps and a bandwidth of 0.2 to 1% [36]. Metal coated (Pt-Ir) tips with a fundamental mechanical resonance frequency of approximately $250\,\mathrm{kHz}$ and tapping amplitude of approximately $100\,\mathrm{nm}$ were used. The detected signals were demodulated at the second harmonic of the tip tapping frequency to suppress background noise. Backscattered signals were collected using either a bolometer or a MCT detector depending on the spectral regime. Given the comparatively low signal-to-noise ratio of the FEL light source due to intensity instabilities, a self-homodyne detection scheme was adopted.

*Finite Element Method Simulations.*

Finite element method (FEM) simulations of the near-field optical response of bulk LAO were performed using the Wave Optics module of COMSOL Multiphysics. The illuminated AFM tip is modelled as a vertically-oriented electric point dipole, placed $2\,\mu\mathrm{m}$ above the LAO surface. To mimick the SNOM measurements, a point probe was placed $1\,\mathrm{nm}$ above the LAO surface and both dipole and probe points were raster-scanned over a $30\,\mu\mathrm{m} \times 30\,\mu\mathrm{m}$ lateral $(x, y)$ area. The amplitude, $|E_z|$, and phase, $\Phi(E_z)$, of the out-of-plane component of the electric field were recorded as a function of the $(x, y)$ dipole and probe positions. For the simulations with a fixed dipole position, a snapshot of the $|E_z|$ field distribution is taken $1\,\mathrm{nm}$ above the LAO sample's surface. We use as an input the dielectric permittivity extracted from the FTIR measurements. In the simulated area, domain "I" has its optical axis aligned along $[111]$, and domains II–IV correspond to successive $90°$ anticlockwise rotations (labels in Fig. 2d).

## Acknowledgements

JT-G acknowledges financial support from the SNSF Swiss Posdoctoral Fellowship (TMPFP2_224378). Parts of this research were carried out at the ELBE Center for High-Power Radiation Sources at the Helmholtz–Zentrum Dresden–Rossendorf e.V., a member of the Helmholtz Association. JW, MR, FGK, RB, SCK, LME, and SDS acknowledge the financial support by the Bundesministerium für Bildung und Forschung (BMBF, Federal Ministry of Education and Research, Germany, Project Grant Nos. 05K19ODB, and 05K22ODA) and by the Deutsche Forschungsgemeinschaft (DFG, German Research Foundation) through Project No. CRC1415 (ID 417590517) and the Würzburg-Dresden Clusters of Excellence ct.qmat – Complexity and Topology in Quantum Matter, and ctd.qmat – Complexity, Topology, and Dynamics in Quantum Matter (EXC 2147, project-id 390858490). This research used resources of the Advanced Light Source, which is a DOE Office of Science User Facility under contract no. DE-AC02-05CH11231. The research of ABK was supported by the Swiss National Science Foundation, Research Grant No. 200021-236697.

## Data availability

All data supporting the findings of this study are available in the Article and the Supplementary Information. Additional data can be obtained from the corresponding authors upon reasonable request.

## Competing interests

The authors declare no competing interests.

## Author Contributions

S.D.S, and J.T.-G. conceived the study. J.W. performed the s-SNOM measurements with the help of J.T.-G J.M.K., and F. G. K.. J.T.-G extracted the dielectric permittivity of LAO with the help of J.T. and D.M.. G.E. performed the Au deposition on the SINS samples. M.R. and R. B. performed pre-characterisation measurements of the samples. J. T.-G and P.L. performed the simulations. J.T.-G measured the SINS spectra with the help of H.A.B. and S.N.G.C.. J.W., J.T.-G, S.C.K., L.M.E, A.B.K, and S.D.S wrote the article with input from all co-authors. All authors contributed to the scientific discussion.

Supplementary Information for:

# Ultra-high THz-field-confinement at $LaAlO_3$ twin walls

J. Wetzel[1,5*], J. Taboada-Gutiérrez[2*a], M. Roeper[1], F.G. Kaps[1], G. Esposito[2], D. Marchese[2], R. Buschbeck[1], P. Lenz[1], J.M. Klopf[3], H. A. Bechtel[4], S. N. Gilbert Corder[4], J. Teyssier[2], S.C. Kehr[1], L.M. Eng[1,5], A. B. Kuzmenko[2], and S.D. Seddon[1a]

[1]Institute of Applied Physics, TUD Dresden University of Technology, Nöthnitzer Straße 61, 01187 Dresden, Germany
[2]University of Geneva, Department of Quantum Matter Physics (DQMP), Geneva 1211, Switzerland
[3]Institute of Radiation Physics, Helmholtz-Zentrum Dresden Rossendorf, 01328 Dresden, Germany
[4]Advanced Light Source (ALS), Lawrence Berkeley National Laboratory, Berkeley, CA 94720, USA
[5]Würzburg-Dresden Cluster of Excellence (EXC2147) ctd.qmat – Complexity, Topology, and Dynamics in Quantum Matter, Dresden, Germany

*These authors contributed equally
[a]email: samuel.seddon@tu-dresden.de, and javier.taboadagutierrez@unige.ch

## **Supplementary Section S1:** Extraction of the IR permittivity of $LaAlO_3$(001)

To extract the infrared (IR) permittivity of $LaAlO_3$ (LAO), we employed a Bruker Vertex v70 Fourier Transform Infrared Spectrometer (FTIR), coupled to a Bruker Hyperion 2000 microscope that allows us to exchange between different objectives. We used a 15x objective with a numerical aperture (NA) of $0.4$, corresponding to an average incidence angle of $\approx 13°$, and measured in reflection mode. In our system, light from a broadband IR source (globar) is modulated by a Michelson interferometer and then focused on the sample by means of a Cassegrain objective. Cassegrain objectives consist of a primary concave parabolic mirror and a secondary convex hyperbolic mirror which make them a versatile tool for broad wavelength applications (such as FTIR) as they are based on reflection of light instead of refraction, avoiding chromatic aberration of light and ensuring a constant focal length of the objective across all IR wavelengths. Then, the reflected light is directed to the detector by the Cassegrain objective again. Furthermore, the microscope allows us to control the polarisation of light both before and after illuminating the sample. In our measurements, a MCT detector for mid-infrared (MIR) frequencies (580 to 8000 $cm^{-1}$) and a bolometer for far-infrared (FIR) frequencies (120 to 700 $cm^{-1}$) were employed. Figure S1 shows a sketch of the whole spectroscopic system, labelling all optical components along the light beam path and highlighting the interferometer, Cassegrain objective and light sources and detector.

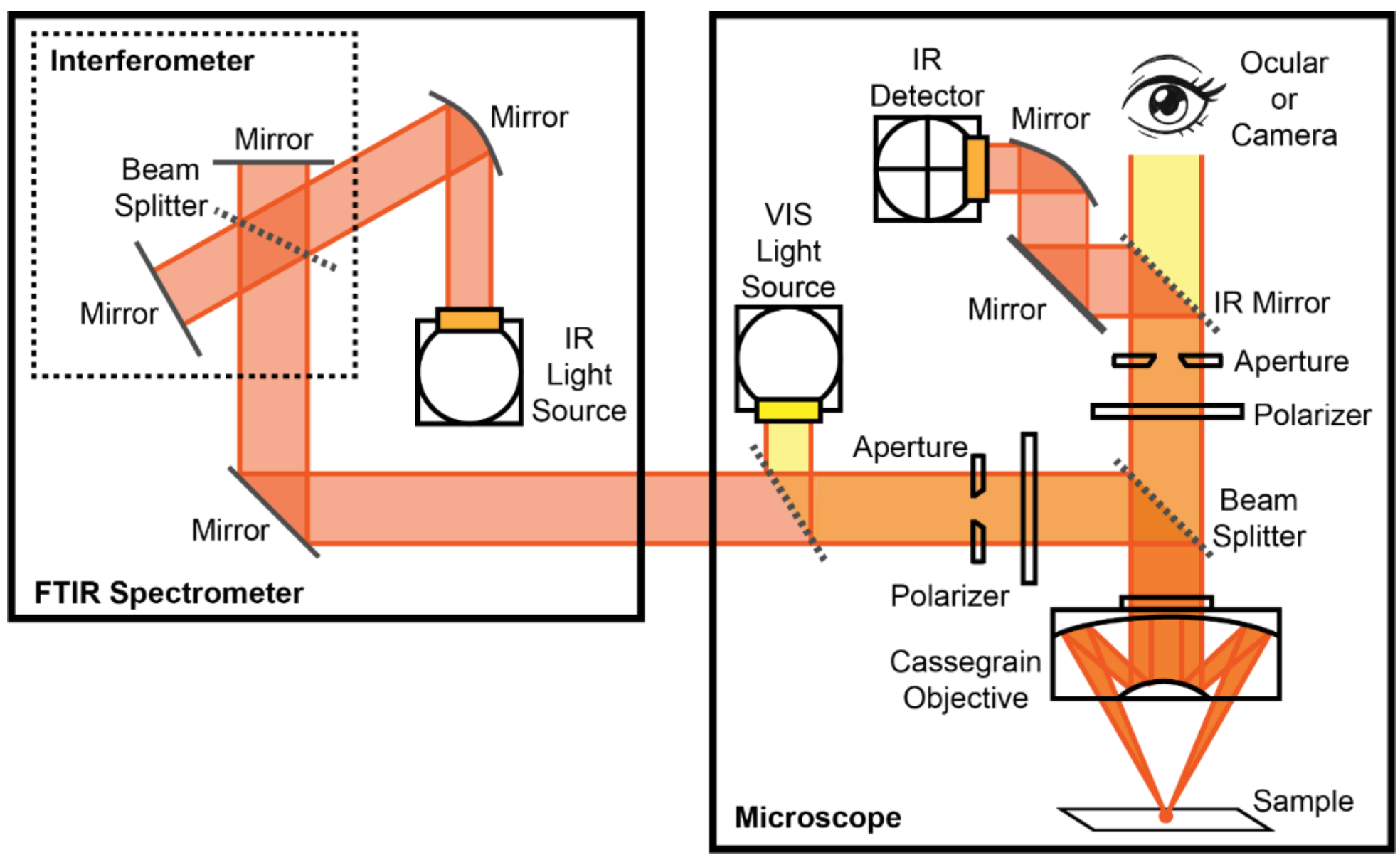

**Figure S1.** Schematics of the FTIR system coupled to a microscope. Light from an IR source (typically a globar) is modulated by means of a Michelson interferometer and then focused onto sample via a Cassegrain objective. The reflected light is collected again by the same Cassegrain objective and directed to the detector (MCT or bolometer) thanks to a set of mirrors. Furthermore, the setup provides us with precise control of light polarisation and incidence angle.

$LaAlO_3$ (LAO) is a distorted perovskite oxide, with R-3c space group, which in its rhombohedral phase, has cell parameters are given by $a = b = c = 5.38552$Å and $\alpha = \beta = \gamma = 60.0109°$. LAO is a wide-bandgap insulator ($E_g \approx 5.6eV$ ), and therefore, in the IR regime, contributions to the optical conductivity are expected only from phonon excitations. In particular, LAO is a uniaxial optical material, which means that its permittivity matrix is given by the following tensor:

$$\hat{\varepsilon} = \begin{pmatrix} \varepsilon_\perp & 0 & 0 \\ 0 & \varepsilon_\perp & 0 \\ 0 & 0 & \varepsilon_\parallel \end{pmatrix},$$

where $\varepsilon_\perp$ is the permittivity component perpendicular to the optical axis and $\varepsilon_\parallel$ the dielectric component parallel to the optical axis. In the case of LAO, the optical axis is oriented with the [111] crystallographic direction for the rhombohedral phase (in the case of the hexagonal representation it would be aligned with the [001] direction). Consequently, the extraction of the full dielectric constant tensor of the material will require to perform polarisation-resolved reflectivity measurements. Ideally, a crystal grown along the [111] direction would simplify measurements and data analysis, but such crystals exhibit twinned domains with random sizes typically ranging from 30 to 60 μm in length, which means that FIR measurements will have undesired contributions to the overall reflectivity coming from adjacent domains, even for the smallest spot sizes achievable with our microscope, which are around 100x100 $\mu m^2$ (limited by diffraction effects in FIR frequencies). Instead, we used a monodomain LAO sample grown along the [110] direction with a polished surface. Nevertheless, in this sample, the optical axis is tilted relative to the surface, resulting in a permittivity tensor given by:

$$\hat{\varepsilon} = \begin{pmatrix} \varepsilon_\perp \cos^2\alpha_{oa} + \varepsilon_\parallel \sin^2\alpha_{oa} & 0 & (\varepsilon_\perp - \varepsilon_\parallel)\cos\alpha_{oa}\sin\alpha_{oa} \\ 0 & \varepsilon_\perp & 0 \\ (\varepsilon_\perp - \varepsilon_\parallel)\cos\alpha_{oa}\sin\alpha_{oa} & 0 & \varepsilon_\parallel \cos^2\alpha_{oa} + \varepsilon_\perp \sin^2\alpha_{oa} \end{pmatrix},$$

with $\alpha_{oa} = 35.1548°$ the angle that forms the direction of the optical axis ([111] direction) with the surface normal ([110] direction), determined from the crystal structure. This tilt complicates permittivity extraction, as the optical axis projection on the XY plane (sample surface) introduces an azimuthal angle $\gamma_{oa}$ which is initially unknown. Figure S2 sketches the geometry of our system, showing the incidence and polarisation angles for the incident light ($\theta_i$ and $\phi_i$, respectively) and the zenith and azimuthal angles for the optical axis ($\alpha_{oa}$ and $\gamma_{oa}$, respectively).

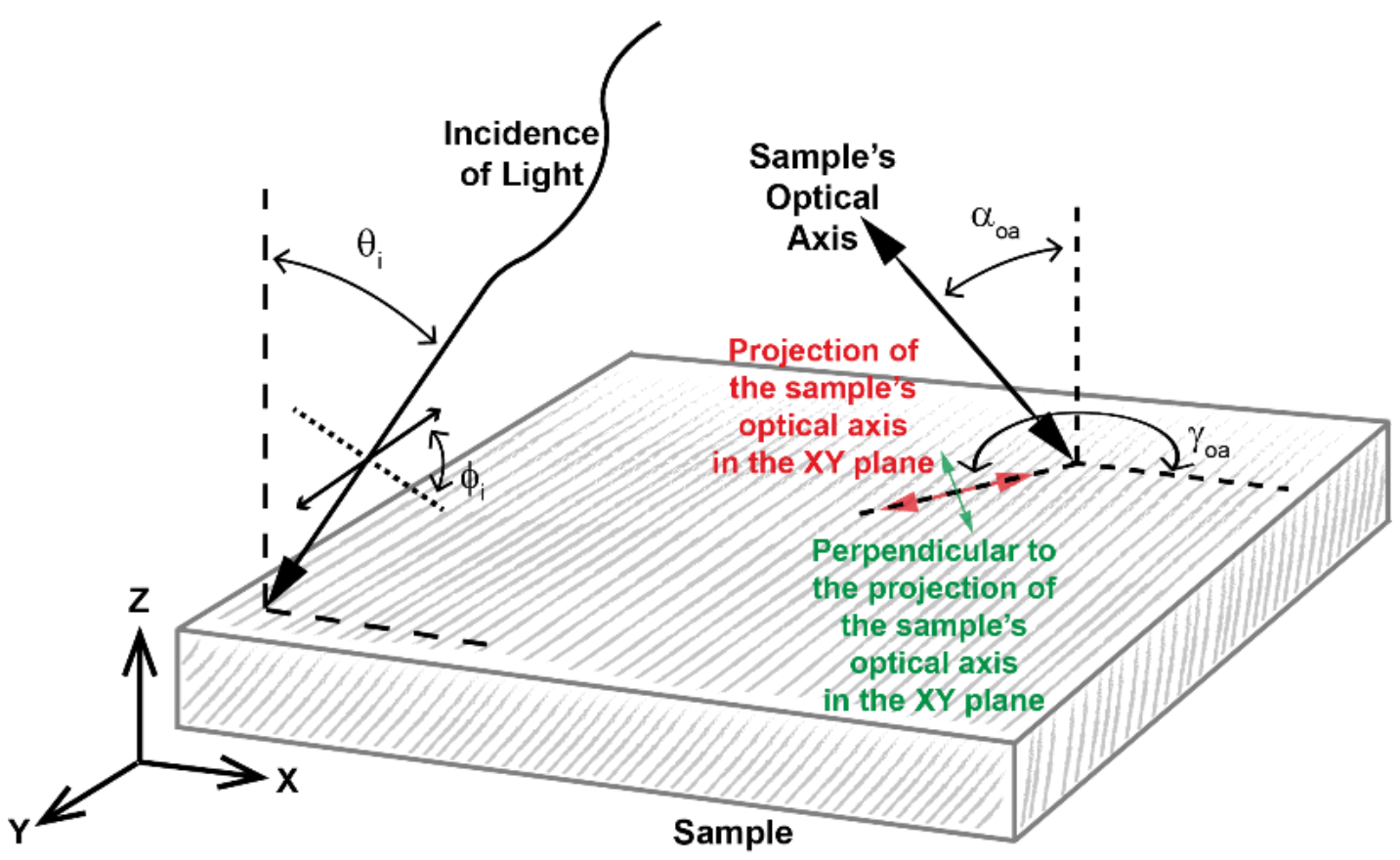

**Figure S2.** Schematics of the incidence of light on our sample. Incident light with angle $\theta_i$ and polarisation angle $\phi_i$ illuminates the sample. The sample is characterized by its tilted uniaxial permittivity tensor, $\hat{\varepsilon}$. The optical axis, tilted at angle $\alpha_{oa} = 35.1548°$ relative to the surface normal, has an azimuthal angle, $\gamma_{oa}$, in the XY plane. The red and green arrows indicate the directions parallel and perpendicular to the optical axis projection, respectively.

This situation, although inconvenient, is not unmanageable. Our first goal is to measure reflectivity along the optical axis projection on the XY plane and its perpendicular direction (stressed by the red and green arrows in the sketch of Figure S2, respectively). Reflectivity measured with the electric field oriented parallel to the optical axis projection on the XY plane (red arrow in Figure S2) will give us a reflectivity that is a function of both $\varepsilon_\perp$ and $\varepsilon_\parallel$, $R_{pr} = R_{pr}(\varepsilon_\perp, \varepsilon_\parallel) = R_{pr}(\varepsilon_{eff})$, due to the tilting of the optical axis, whereas reflectivity measured for the electric field oriented along the direction perpendicular to that projection will give us directly a reflectively which is purely coming from the $\varepsilon_\perp$ component, $R_\perp = R_\perp(\varepsilon_\perp)$. Since the azimuthal angle of the optical axis $\gamma_{oa}$ is unknown and we cannot align the polarisation of light with the optical axis projection, we measured reflectivity at four random polarisation angles $\phi_i$ we called 0°, 45°, 90° and −45°, as the reflectivity for any polarisation angle $\phi_0$ can be calculated from any set of 3 independent reflectivity measurements [1]:

$$R(\phi_0, \omega) = aR(\phi_1, \omega) + bR(\phi_2, \omega) + cR(\phi_3, \omega) \qquad \text{(Eq. S1)}$$

With $\phi_1$, $\phi_2$ and $\phi_3$ the polarisation angles for each measurement, $\gamma$ the polarisation angle at which we want to extract the reflectivity spectrum and $a$, $b$ and $c$ given by:

$$a = \frac{\sin(\phi - \phi_2)\sin(\phi - \phi_3)}{\sin(\phi_1 - \phi_2)\sin(\phi_1 - \phi_3)}$$

$$b = \frac{\sin(\phi - \phi_1)\sin(\phi - \phi_3)}{\sin(\phi_2 - \phi_1)\sin(\phi_2 - \phi_3)}$$

$$c = \frac{\sin(\phi - \phi_1)\sin(\phi - \phi_2)}{\sin(\phi_3 - \phi_1)\sin(\phi_3 - \phi_2)},$$

Figure S3 shows the reflectivity spectra at these four polarisations, revealing three high-reflectivity bands associated with phonon modes (the so-called Reststrahlen bands, RBs) between approximately 150 and 280 cm$^{-1}$, 430 and 600 cm$^{-1}$, and 650 and 780 cm$^{-1}$. The largest differences occur near 592 cm$^{-1}$ and in the third RB.

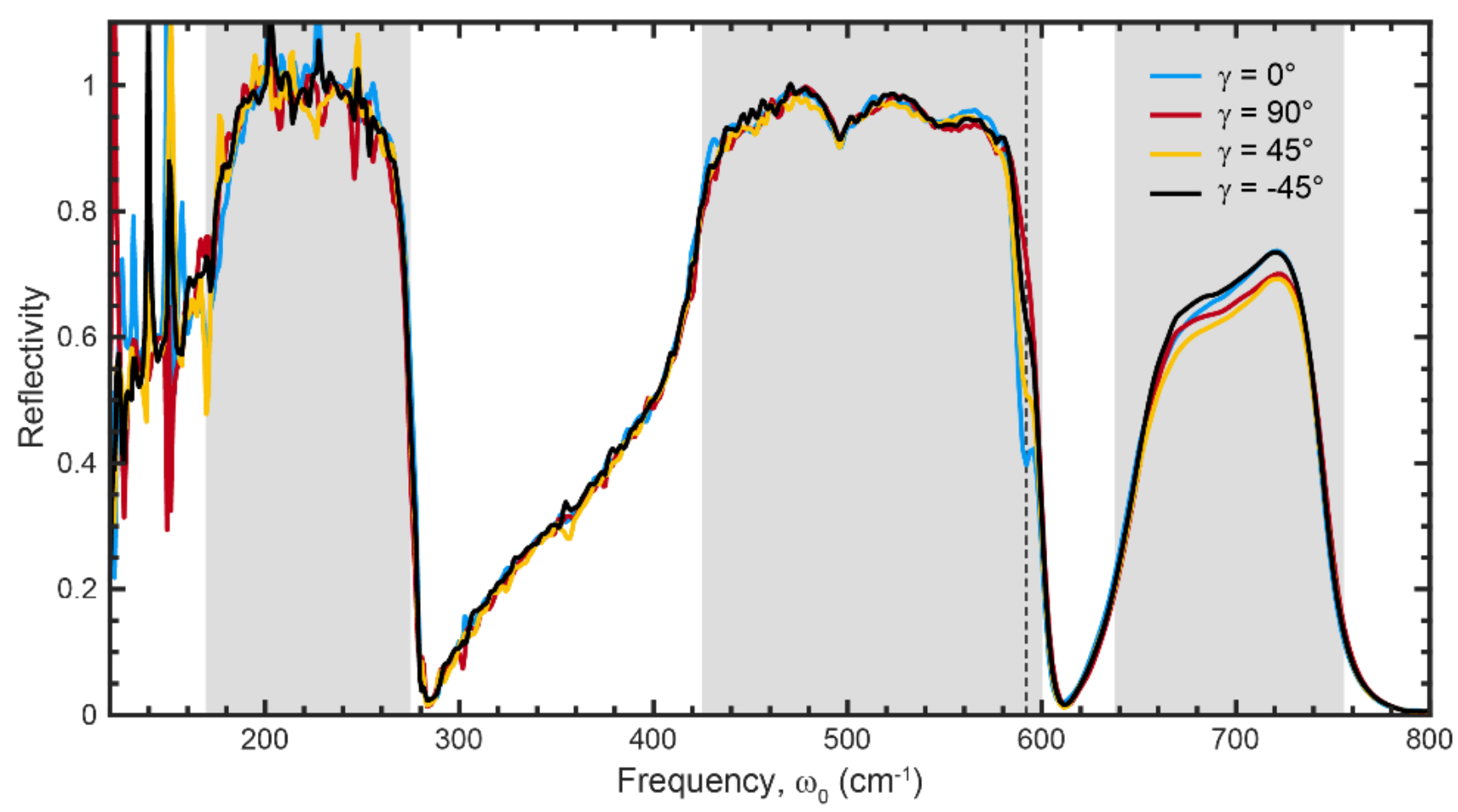

**Figure S3.** IR reflectivity spectra of LAO at four polarisation angles. Spectra at 0° (blue curve), 90° (red line), 45° (yellow curve) and $-45°$ (black line) identify 3 RBs, stressed by grey shaded regions, a first one between ≈ 150 and 280 cm$^{-1}$, a second one from ≈ 430 to 600 cm$^{-1}$, and a third one between ≈ 640 and 760 cm$^{-1}$. We observe the largest differences between all polarisations close to the $\omega_{LO}$ phonon at ≈ 592 cm$^{-1}$ (vertical dashed line) and in the third RB.

To verify Equation S1, we have intentionally used the spectra at $\phi_i = 0°$, $90°$ and $45°$ to recover the reflectivity spectrum at $\phi_i = -45°$. Figure S4 compares the calculated and measured spectra, showing excellent agreement.

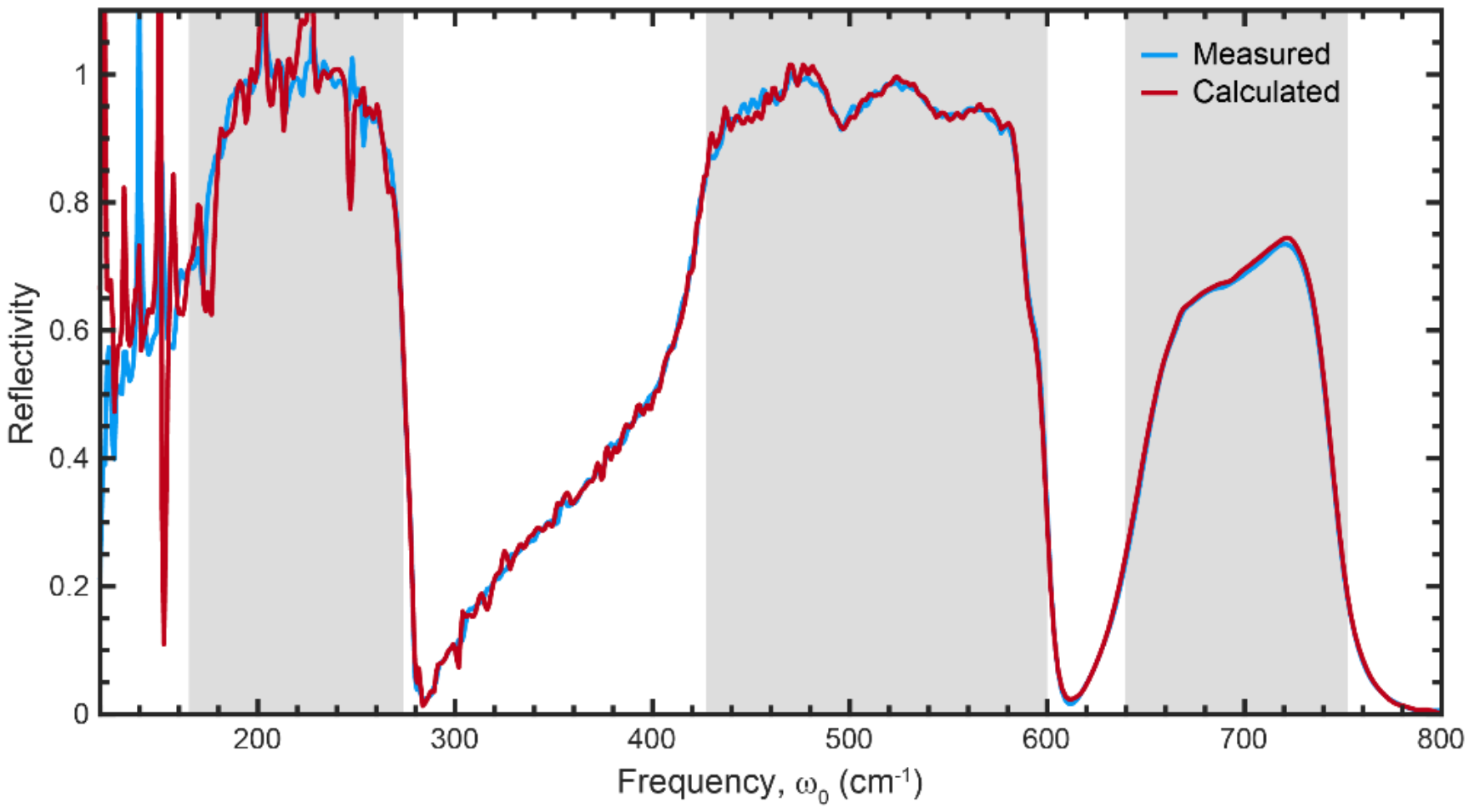

**Figure S4.** Validation of polarisation reconstruction. The LAO measured reflectivity spectrum at $\phi_i = -45°$ is compared with the spectrum calculated employing Equation S1 and the spectra at $\phi_i = 0°, 90°$ and $45°$. The calculated spectrum (red line) is fully in line with the measured one (blue curve). RBs are stressed by grey shaded regions.

For a certain frequency, the largest reflectivity spectral contrast occurs between polarisations aligned along and across the optical axis projection and therefore, employing Equation S1 and scanning for all possible polarisation angles at a given frequency, we can identify at which polarisation angles the difference becomes maximum. We have selected $\omega_0 = 592\ cm^{-1}$ (vertical dashed line in Figure S3) as the difference in reflectivity is maximum between all measured polarisations, see Figure S5.

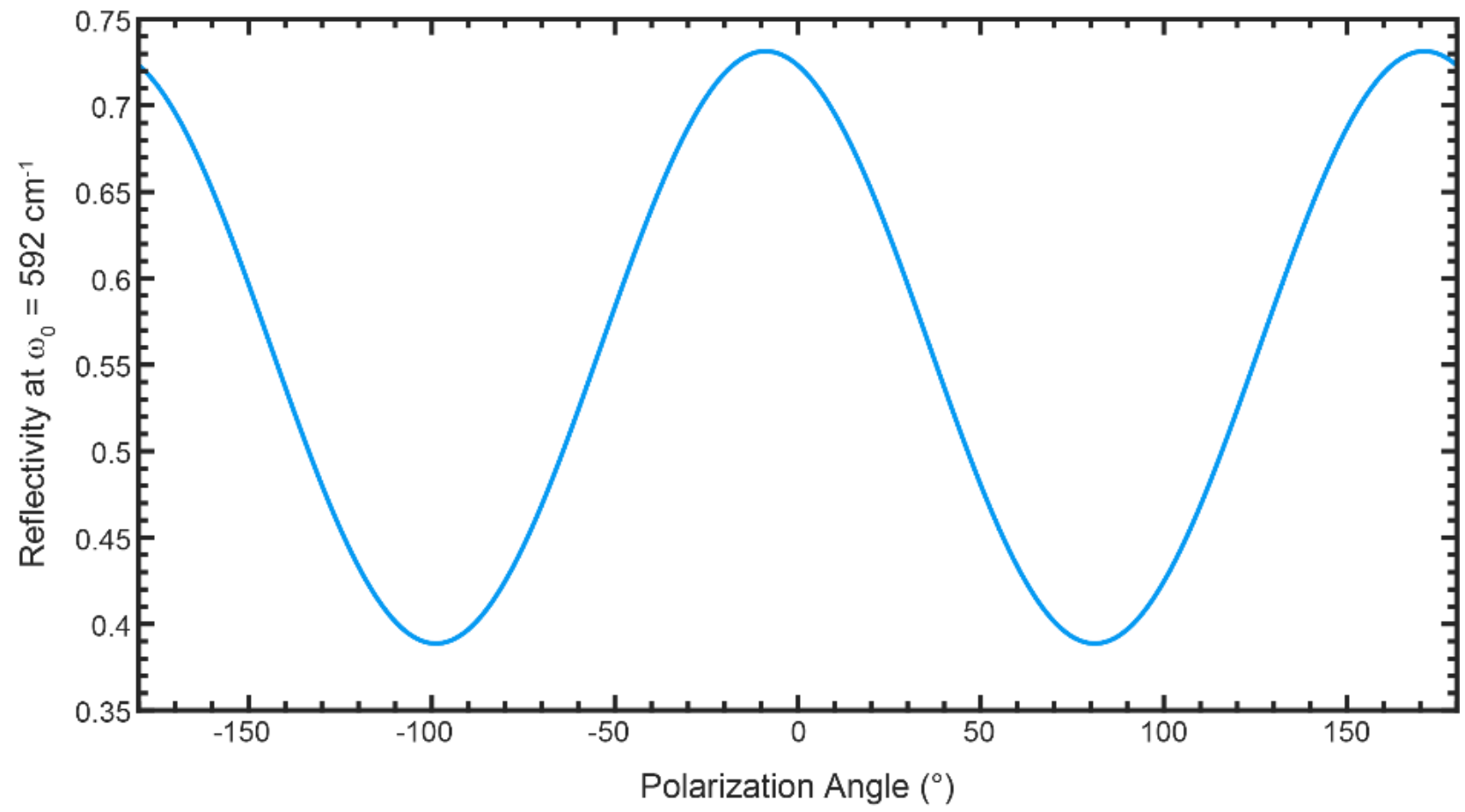

**Figure S5.** Reflectivity at $\omega_0 = 592\ cm^{-1}$ as a function of polarisation angle, showing sinusoidal behaviour with a 180° periodicity. From the minimum ($\phi_i = 81°$) and maximum ($\phi_i = -9°$) positions, we can calculate the reflectivity spectra for the directions parallel and perpendicular to the optical axis projection in the XY plane.

We find a sinusoidal curve with a periodicity equal to 180°, with a maximum at $\phi_i = -9°$ and a minimum at $\phi_i = 81°$. Consequently, we have calculated the whole IR reflectivity spectra employing Equation S1 for those 2 polarisation angles, shown in Figure S6. The spectrum at $\phi_i = 81°$ (red curve) reflects a coupling between the two permittivity components $\varepsilon_\perp$ and $\varepsilon_\parallel$ ($R_{min} = R_{min}(\varepsilon_\perp, \varepsilon_\parallel) = R_{min}(\varepsilon_{eff})$,) that can be identified thanks to the kink which appears close to the edge of the second RB at around 590 cm$^{-1}$. If that reflectivity would have come from a pure dielectric component, it would not have shown any kink, and the spectra would have decayed stepless as it is indeed shown in the other curve that corresponds to $\phi_i = -9°$ and that we can associate with $R_{max} = R_\perp(\varepsilon_\perp)$.

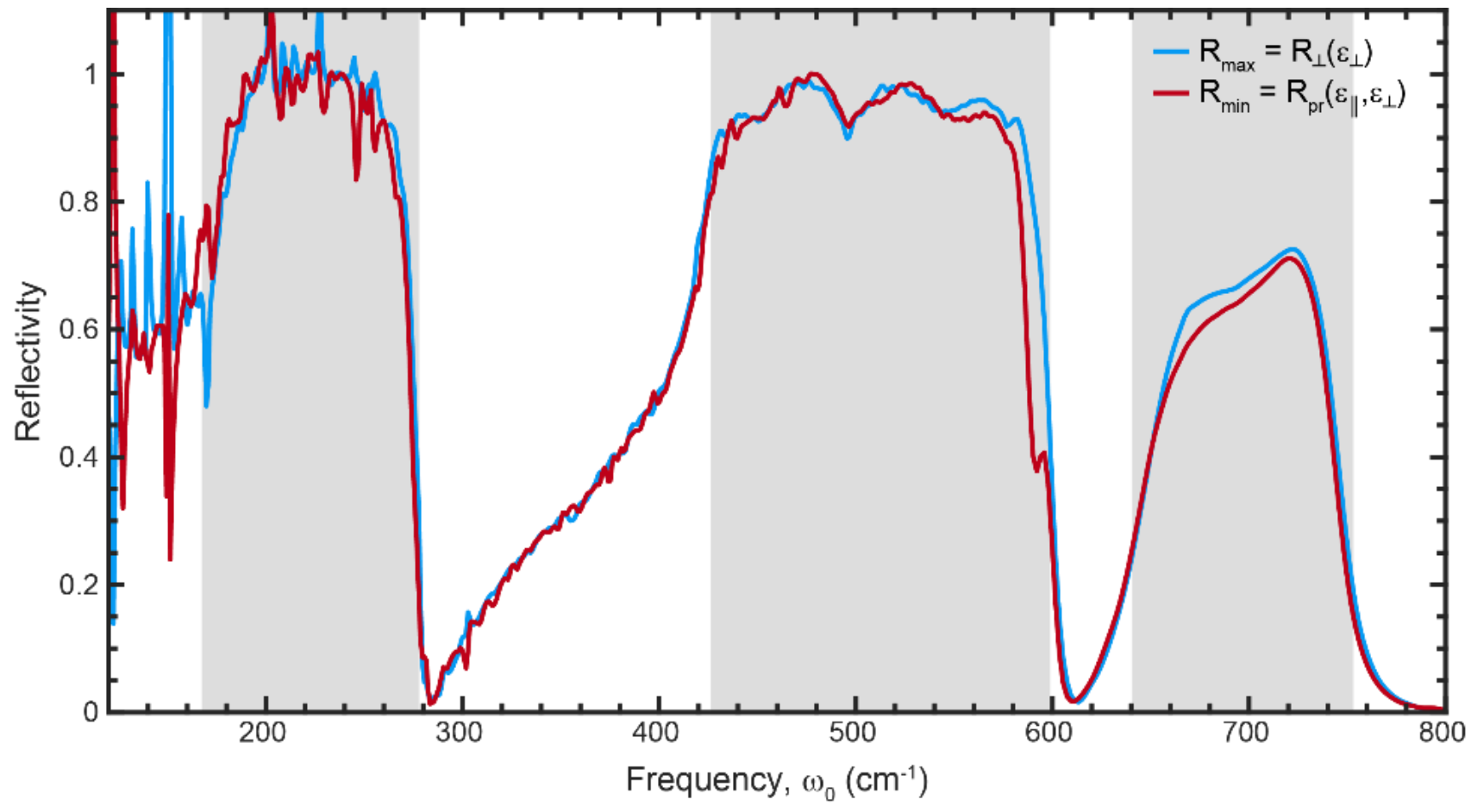

**Figure S6.** Reflectivity spectra at maximum and minimum polarisation angles obtained in Figure S5. The reflectivity spectrum at $\phi_i = 81°$ (red curve, $R_{min} = R_{pr}(\varepsilon_\parallel, \varepsilon_\perp)$) shows a mixed contribution between $\varepsilon_\parallel$ and $\varepsilon_\perp$, while the spectrum calculated at $\phi_i = -9°$ (blue curve, $R_{max} = R_\perp(\varepsilon_\perp)$) shows a pure contribution from $\varepsilon_\perp$. RBs are depicted by grey shaded regions.

We can directly fit $R_{max}$ to extract $\varepsilon_{\perp}$. $\varepsilon_{\parallel}$ on the other hand, can be extracted by supposing that the optical response of the $R_{min} = R_{min}(\varepsilon_{mix})$ curve is given by an effective permittivity that mixes both $\varepsilon_{\parallel}$ and $\varepsilon_{\perp}$ [2]:

$$\varepsilon_{mix} = \varepsilon_{\infty,mix} + \frac{(\varepsilon_{\perp}-\varepsilon_{\infty,\perp})\cos^2\alpha_{oa}+(\varepsilon_{\parallel}-\varepsilon_{\infty,\parallel})\sin^2\alpha_{oa}+(\varepsilon_{\parallel}-\varepsilon_{\infty,\parallel})(\varepsilon_{\perp}-\varepsilon_{\infty,\perp})}{1+(\varepsilon_{\perp}-\varepsilon_{\infty,\perp})\sin^2\alpha_{oa}+(\varepsilon_{\parallel}-\varepsilon_{\infty,\parallel})\cos^2\alpha_{oa}}, \qquad \text{(Eq. S2)}$$

With $\varepsilon_{\infty,mix}$, $\varepsilon_{\infty,\parallel}$, $\varepsilon_{\infty,\perp}$ the high-frequency permittivities for the “effective mixed” state, parallel and perpendicular components of the dielectric tensor. We fitted the reflectivity spectra employing RefFit, a software for fitting optical spectra with a wide variety of dielectric-function models [3], setting $\alpha_{oa} = 35°$, which is the crystallographic angle that the [110] and [111] directions form in LAO. Both $\varepsilon_{\perp}$ and $\varepsilon_{\parallel}$ have been modelled employing a Drude-Lorentz oscillators model:

$$\varepsilon_i(\omega) = \varepsilon_{\infty,i} + \sum_n \frac{\omega_{p,i,n}^2}{\omega_{TO,i,n}^2-\omega^2-i\gamma_{i,n}\omega},$$

With $i = \perp$ or $\parallel$, $\varepsilon_{\infty,i}$ the high-frequency permittivity for the $i$ component of the dielectric tensor and $\omega_{TO,i,n}$, $\omega_{p,i,n}$ and $\gamma_{i,n}$ the transversal optical frequency, “plasma” frequency and damping factor of the $n$-th mode, respectively. We have employed 5 modes in both cases. Figure S7 shows the fittings to the reflectivity $R_{max}$ and $R_{min}$ spectra. Table S1 summarizes the fitted parameters for both $\varepsilon_{\parallel}$ and $\varepsilon_{\perp}$. For simplicity, and because the pseudocubic nature of the material leads to only a small frequency shift between the $\varepsilon_{\parallel}$ and $\varepsilon_{\perp}$ dielectric components, we refer to the Reststrahlen bands as RB1, RB2, and RB3. Strictly speaking, however, one should further differentiate between the permittivity components.

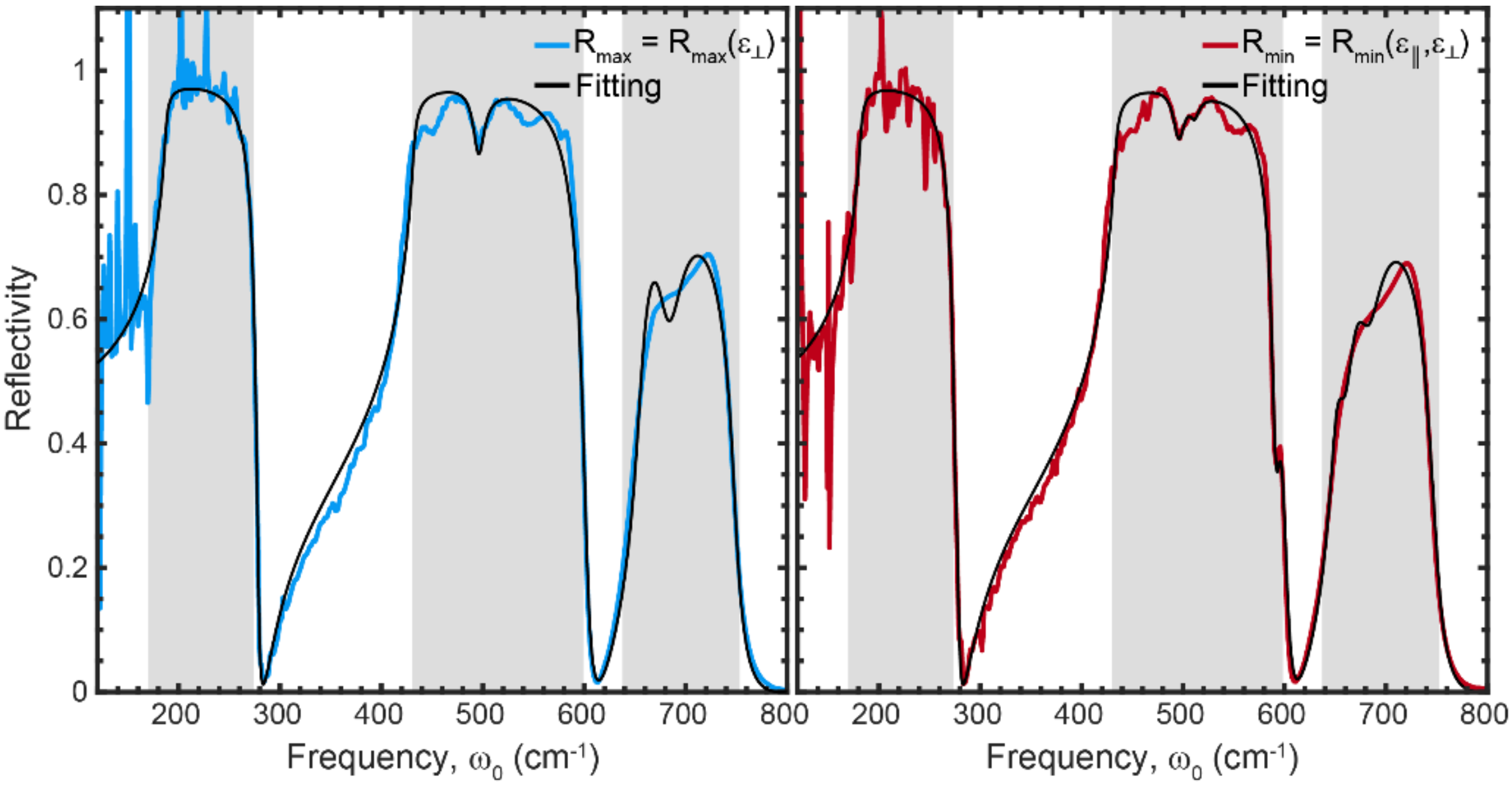

**Figure S7.** Fitting the reflectivity spectra. The left panel represents $R_{max} = R_{\perp}(\varepsilon_{\perp})$, blue line, and the fitted curve, black line, supposing normal incidence of light. Right panel represents $R_{min} = R_{pr}(\varepsilon_{\parallel}, \varepsilon_{\perp})$, red curve and the fitted curve, black line, employing Equation S2. $R_{max}$ has been fitted employing a simple Drude-Lorentz oscillators model, with 5 phonon modes. $R_{min}$ has been fitted with input from $\varepsilon_{\perp}$ fitted in $R_{max}$, employing in addition the 5 phonon modes of $\varepsilon_{\parallel}$.

| Component | $\varepsilon_\infty$ | $\omega_{TO}\ (cm^{-1})$ | $\omega_p\ (cm^{-1})$ | $\gamma\ (cm^{-1})$ |
|---|---|---|---|---|
| $\varepsilon_\perp$ | 5.07 | 187.58 | 779.49 | 4.0124 |
| | | 431.76 | 972.87 | 5.9136 |
| | | 496.79 | 134.33 | 13.602 |
| | | 655.28 | 371.15 | 14.718 |
| | | 684.9 | 160.93 | 29.64 |
| $\varepsilon_\parallel$ | 4.42 | 165.1 | 737.78 | 5.6217 |
| | | 438.27 | 857.88 | 3.4048 |
| | | 512.35 | 94.214 | 10.956 |
| | | 639.91 | 286.09 | 8.8241 |
| | | 672.47 | 277.66 | 26.178 |

**Table S1.** Fitted Drude-Lorentz parameters for $\varepsilon_\parallel$ and $\varepsilon_\perp$.

Figure S8 plots the real part of the modelled permittivity in the three RBs employing the parameters of Table S1, with the red curve representing the real part of the $\varepsilon_\parallel$ permittivity and the blue curve the $\varepsilon_\perp$ component.

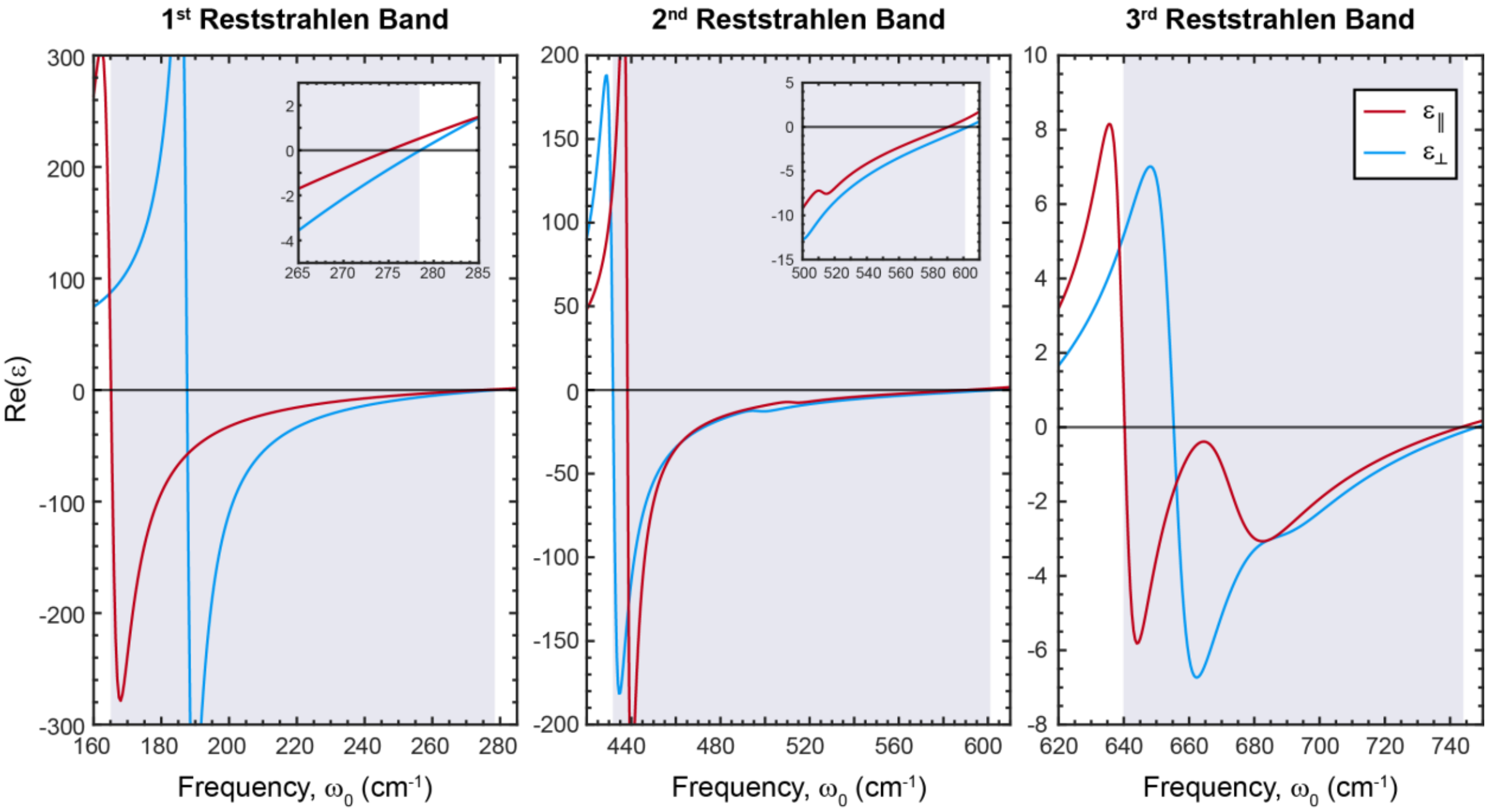

**Figure S8.** Infrared permittivity of LAO. For the sake of clarity, the image has been divided into three panels with different scales. Red curves represent $\varepsilon_\parallel$ while blue curves represent $\varepsilon_\perp$. Grey shaded regions represent the RBs. Inset panels in the 1st and 2nd RBs represent a zoom-in of $\mathrm{Re}(\varepsilon)$ at frequencies close to $\omega_{LO}$, stressing the hyperbolic regimes in the material.

**Supplementary Section S2:** LAO phonon-polariton excitation demonstrated by synchrotron infrared nano-spectroscopy (SINS)

Synchrotron infrared nano-spectroscopy (SINS) was performed in a (001)-grown LAO sample to demonstrate polaritonic launching and excitation. SINS is a spectroscopic technique which is based on a scattering-type scanning near-field optical microscope (s-SNOM), which allows nanoscale resolved absorption spectra of the samples to be obtained with wavelength-independent spatial resolution. Figure S9A shows an optical microscope image of the LAO sample employed for demonstrating the polaritonic launching via a gold bar fabricated through optical lithography. The picture has been taken with an optical microscope employing 2 crossed polarizers for the sake of visualization of the LAO twin domains, implying that light reflected from the Au bar is blocked, and thus its black colour. Figure S9B shows a SINS point spectrum (third harmonic of the tip oscillation frequency, normalised to Au, indicated by the yellow dot in Figure S9A) taken at the centre of a twin domain, far from any twin boundary. The spectrum shows an enhancement of the near-field optical signal at frequencies within the three RBs of LAO, indicating polaritonic excitation within these frequency bands. To demonstrate the polaritonic launching and measure the dispersion, we have performed SINS linescans at a direction perpendicular to the gold bar (along yellow dashed line in figure S9A, scaled to the total length of the SINS linescan). Figure S9C shows the third harmonic SINS linescan spectra normalised to the Au reference, $s_3/s_{3,Au}$. We can clearly see maxima of the near-field signal at frequency regions coinciding with the LAO's RBs, in line with the spectrum shown in Figure S9B. Furthermore, we observe signal oscillations within the first and second RBs (the signal is too low in the 3rd RB of LAO as to observe any oscillation, due to the intrinsic losses of the material at these frequencies), with an oscillation period that decreases as the light's frequency increases. Figure S9D plots some of the SINS spectra at frequencies that clearly show the fringe oscillations. We have artificially shifted them along the vertical direction for the sake of clarity. We can assign these oscillations with the propagating surface phonon polaritons excited by the Au bar [4]. By fitting the spectral profiles to an exponentially decaying sinusoidal function, we can extract the polaritonic wavelength:

$$y = y_0 + A \cdot \exp\left(-x/L_p\right) \cdot \sin(2\pi x/\rho + \phi_0), \qquad \text{(Eq. S3)}$$

with $y_0$ an offset to the near-field signal, $A$ the amplitude of the polaritonic wave, $L_p$ the propagation length, $\rho$ the fringe spacing and $\phi_0$ a phase offset. It should be noted that, as we are working with bulk LAO, polaritons are not highly confined, and so the periodicity of the measured near-field oscillations (the fringe spacing, $\rho$) does not correspond to the polaritonic wavelength, $\lambda_p$ [4]. By geometrical considerations, it can be calculated that the fringe spacing $\rho$ can be related to the polaritonic wavelength, $\lambda_p$, by means of the following equation [5, 6]:

$$\lambda_p = \frac{\rho}{1+\frac{\rho}{\lambda_0}\cos\alpha}, \qquad \text{(Eq. S4)}$$

with $\lambda_0$ the wavelength of the incident light, $\lambda_0 = 1/\omega_0$, and $\alpha$ the angle of the incident light, $\alpha \approx 45°$. It should be noted that, due to the transition from one twin domain to the adjacent one during the SINS linescan (see yellow dashed line in Figure S9a) polaritons are weakly reflected back at the domain wall. This is not an issue for extracting $\lambda_p$ due to the very small difference between the permittivity components $\varepsilon_\perp$ and $\varepsilon_\parallel$, differing basically in the optical axis orientation. However, it clearly affects the damping term (the propagation length $L_p$), extremely sensitive to inhomogeneities at the surface of the polaritonic material. Therefore, no reliable experimental information about the polaritonic propagation length or lifetime, $\tau$, can be extracted.

Employing the permittivity calculated in Supplementary Section S1, we have calculated the imaginary part of the Fresnel reflection coefficient as a function of momentum and frequency, $r_{pp}(q,\omega)$, through a transfer matrix method as the polaritonic dispersion is contained within the denominator of $r_{pp}(q,\omega)$ [7]. The calculation is shown in Figure S9e as a false colour plot, confirming a polaritonic branch in every RB. The experimental dispersion, calculated by fitting the SINS profiles with Equations S3 and S4, with the experimental polaritonic momentum given by $q_p = 2\pi/\lambda_p$, is represented by the blue dots, obtaining excellent agreement between measurements and calculations.

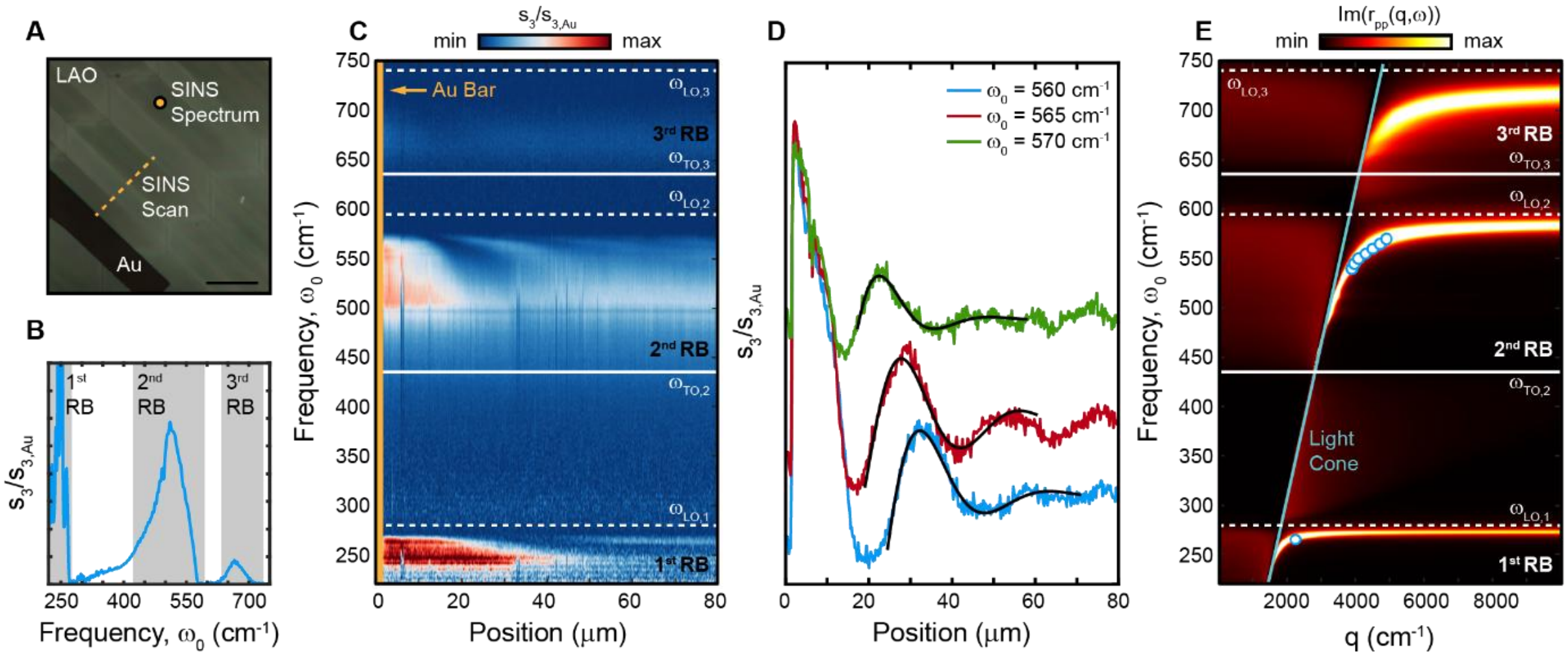

**Figure S9.** Polaritonic launching in LAO through a gold bar. **A** Optical image of a [100] LAO substrate presenting its typical chevron pattern. An Au rod was fabricated as polaritonic launcher by means of optical lithography. 2 crossed polarizers were employed for taking the picture, as it enhances the optical contrast between the different LAO domains. Scale bar is 50 microns. **B** LAO SINS spectrum (third harmonic, normalised to gold) taken at the yellow point position of **A**. We clearly observe an enhancement of the near-field signal at the frequency regions within the RBs. **C** LAO SINS linescan taken along the yellow dashed line in **A**. Polaritonic oscillations can be clearly observed within the first and second RBs. The Au bar position is marked with a yellow stripe at the very left of the image. **D** Near field $s_3/s_{3,Au}$ profiles extracted from **C** at selected frequencies (560, 565, and 570 cm$^{-1}$ in blue, red, and green, respectively) showing the polaritonic oscillations. The profiles were fitted employing Equation S3 for extracting fringe periodicity. Black lines represent the fittings for every profile. **E** LAO polaritonic dispersion (false colour plot) calculated through the transfer matrix method. We can observe a polaritonic branch within every RB. The dispersion of light is depicted as a light blue solid line. Blue circumferences depict the polaritonic dispersion measured by extracting the polaritonic wavelength $\lambda_p$ from the profiles in **D** employing Equations S3 and S4.

**Supplementary Section S3:** Twin wall junctions in LAO and their near-field signal in all Reststrahlen bands measured by FEL-SNOM

In the main text the twin wall (TW) notation was introduced with subscripts v/d to differentiate between vertical and diagonal TWs, and a superscript of $\parallel/\perp$ to indicate the frequency-dependent optical behaviour of each set of TWs. Since the designations *vertical* and *diagonal* are contingent on sample orientation and carry no crystallographic meaning, we propose a Miller index-based notation in which $TW_{\mathrm{v}}$ and $TW_{\mathrm{d}}$ are replaced by $TW_{\mathrm{hkl}}$, where hkl is trivially the direction along which the TW points, such as (010). Note that given that the TWs along $(1\bar{1}0)$ and $(110)$ within one set of 45° walls are all equivalent in behaviour, for ease these TWs will be described as a set of directions $TW_{\{110\}}$, rather than with their individual Miller indices. The superscript $\parallel$ $(\perp)$ will be the same as in the main text, leading to a total of 4 possible TWs with 2 distinguishable cases: i) $TW^{\parallel}_{(010)}$ and $TW^{\parallel}_{\{110\}}$, (or together simply just $TW^{\parallel}_{(010),\ \{110\}} = TW^{\parallel}$) where the optical axes point toward the TW surface junction; and ii) $TW^{\perp}_{(010)}$ and $TW^{\perp}_{\{110\}}$, (and similarly together denoted as $TW^{\perp}_{(010),\ \{110\}} = TW^{\perp}$) where they point towards the TW, but away from the surface. With always four types of TWs meeting at any twin wall junction (TWJ), Tab. S2 presents the composition of the two possible TWJ:

| | $TW^{\parallel}_{(010)}$ | $TW^{\parallel}_{\{110\}}$ | $TW^{\perp}_{(010)}$ | $TW^{\perp}_{\{110\}}$ |
|---|---|---|---|---|
| $TWJ_1$ | 1x | - | 1x | 2x |
| $TWJ_2$ | 1x | 2x | 1x | - |

**Table S2.** Composition of the two possible TWJs on a LAO(001) surface. Each TWJ consists of in total 4 TWs, 2 oriented along the (010)-plane and 2 along the set of {110}-planes. The number shows whether and how often a TW is present in the respective TWJ.

As shown in Fig. S8, there are 3 Reststrahlen bands of LAO in the THz regime and FEL-s-SNOM measurements were carried out in all three of them, even though only the results of $RB_1$ are shown in the main text, due to the exceptional contrast displayed. The range in which one can visibly differentiate between the bulk material and the TWs in the first RB is from 7.79 THz to 8.16 THz, in the second RB between 16.19 THz and 17.69 THz and in the third RB from 18.59 THz to 20.98 THz. In Fig. S10 measurements of $TWJ_1$ at two specific frequencies are presented for each RB, showing a clear inversion of contrast for $RB_1$ and $RB_2$. At $RB_3$, the contrast is not as high, due to the intrinsic losses of the material in this band, as also found in the SINS data (see S2 and Fig. S9). Notably, qualitatively the same contrast is reproduced within all RBs, demonstrating the resonant character of the TW signal.

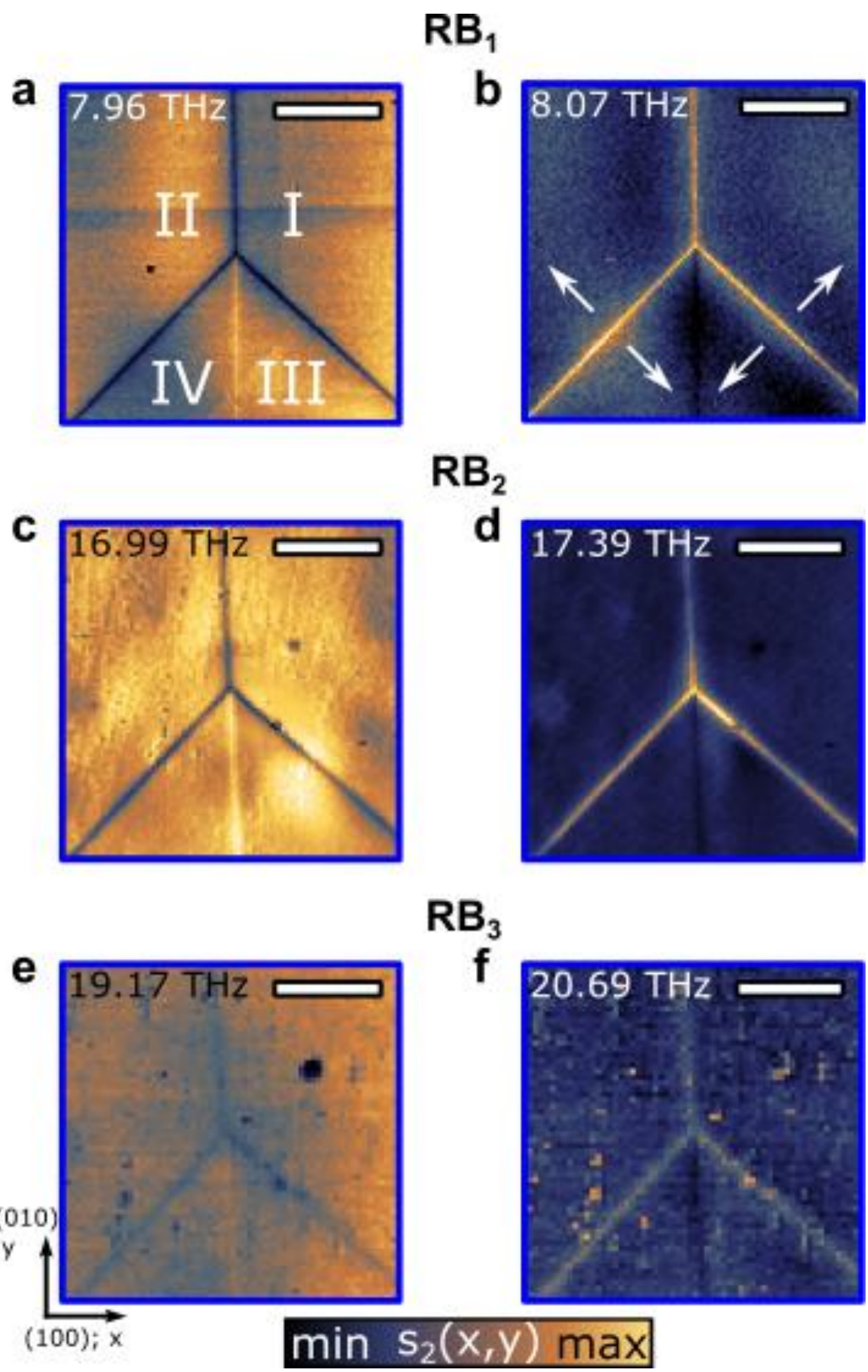

**Figure S10.** Spectral response in all 3 RBs at selected frequencies measured by FEL-SNOM on junction $TWJ_1$. Second-harmonic near-field signal $s_2(x,y)$ **a, b** in the first RB at 7.96 THz and 8.07 THz, **c, d** in the second RB at 16.99 THz and 17.39 THz, and **e, f** s2(x,y) in the third RB at 19.17 THz and 20.69 THz. All scale bars are 10 µm.

At 8.07 THz a high-resolution scan was performed in order to investigate the width of the TWs. The result of that can be seen in Fig. S11. With an excitation wavelength of 37.15 µm (8.07 THz) and a FWHM of 143 nm of $TW^{\perp}_{\{110\}}$ , a lateral confinement of 260 is achieved. Note that the experimental confinement is a convolution of the TW width and the SNOM-tip radius. FEM simulations predict a confinement ration of over 1000 (see main text).

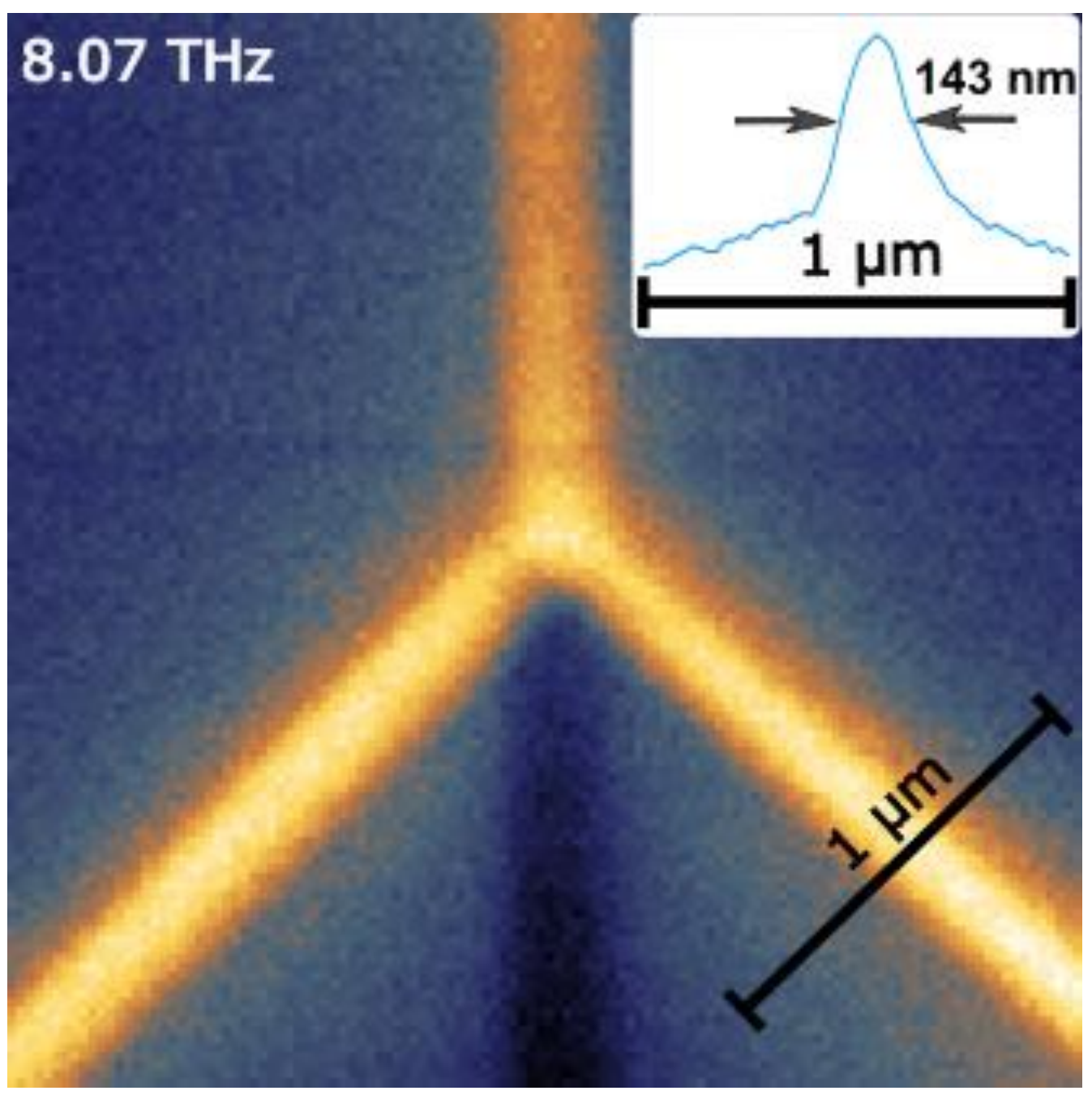

**Figure S11.** High-resolution optical image of $TWJ_1$ at 8.07 THz with the inset showing a line profile of a 1 µm length over $TW^{\perp}_{\{110\}}$ with a FWHM of 143nm, demonstrating a lateral confinement of 260.

## **Supplementary Section S4:** Analytical Point-Dipole Model for Twin-Domain Walls

In order to analytically describe the fundamental interaction of a s-SNOM probe with the LAO twin-domain walls (TW), we modify here the analytical point-dipole model as originally introduced by Knoll and Keilmann [8] and thereby extend it for interaction with an anisotropic sample that includes a domain wall, i.e. a boundary between sample areas with different permittivity tensors. The two adjacent sample domains are therefore represented by two laterally separated polarizable spheres with in general arbitrarily oriented anisotropic permittivity tensors. Similar to the conventional dipole-model, the tip is assumed to be represented by an isotropic sphere and is placed here directly above the sample junction, i.e. symmetrically between the two sample spheres (see Fig. S12A).

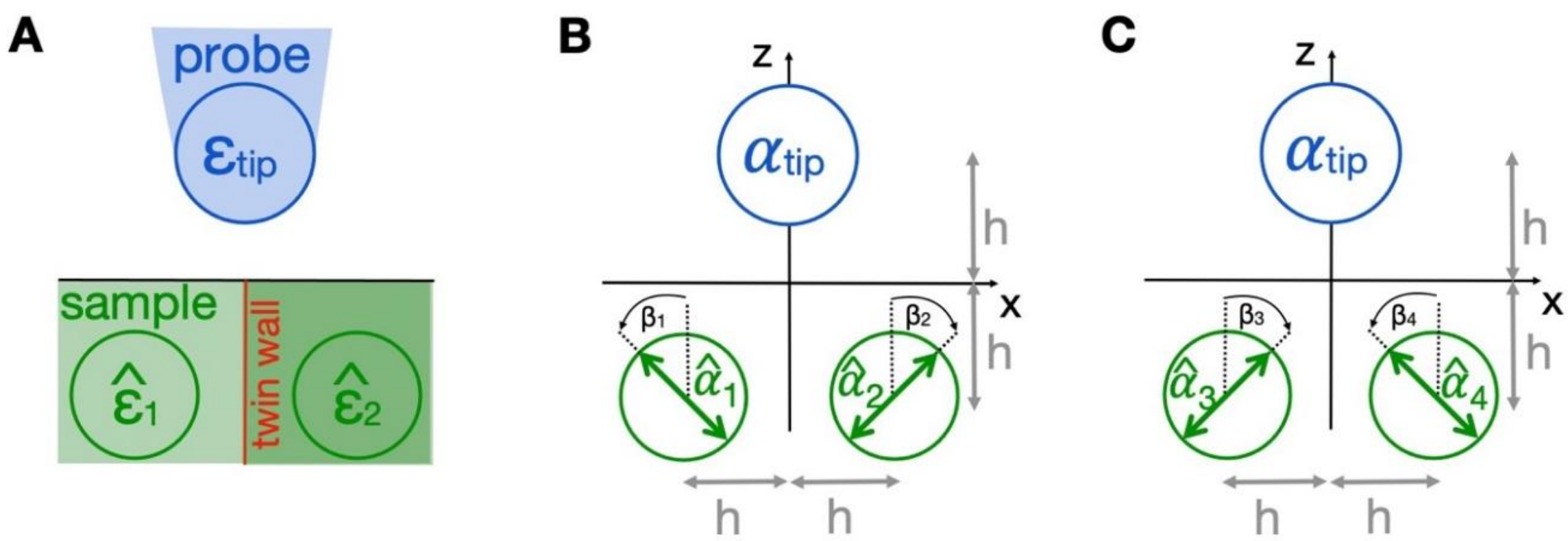

**Figure S12.** Model system representing a near-field probe next to a twin-domain wall (TW) separating two sample areas with different permittivity tensors. **A** General concept with probe and sample areas being represented by polarizable spheres of permittivites $\varepsilon_{tip}$ (isotropic tip) and $\hat{\varepsilon}_1$, $\hat{\varepsilon}_2$ (anisotropic sample areas), respectively. **B,C** The two cases A, B with different orientations of the optical axis (green arrows) and corresponding polarizability tensors $\alpha_i$ at the TWs as discussed in detail in this supplementary (see description in text).

It shall be noted, that this model system acts as a very first approximation only, illustrating the coupling mechanism and the relevant system parameters. Even though it qualitatively reproduces the experimentally and numerically observed response surprisingly well (see Fig. 3 in main text), it is expected to be strongly limited in terms of giving quantitative numbers for e.g. distance dependence and domain/domain-wall contrast. Moreover, it is suited only for tip positions directly above the sample junction as well as for planar domain walls orthogonal to the sample surface. For a more detailed theoretical analysis either numerical simulations (as carried out in this work) or a representation of the sample by a manifold of point dipoles (as applied e.g. in [9]) are needed.

Besides these limitations, the advantages of the here discussed simple extension of the dipole model for qualitative description of the LAO-TW signature include

- the possibility to introduce two sample areas with different optical response,
- direct implementability of arbitrary sample anisotropy (including in particular optical axis non-orthogonal to the sample surface, unlike e.g. in [10, 11]), and
- application to arbitrary excitation polarisation, which are not possible when e.g. using the finite dipole model [12].

The derived model results in a simple, analytical solution of the scattering cross sections in s-SNOM for different anisotropic domain configurations. Assuming twin-domain walls with mirrored symmetry, the model can be solved fully analytically resulting in formulas with distinct similarities compared to the conventional point-dipole model as shown below. In the following, without restriction, we exemplarily show the derivation of these formulas for excitation fields strictly orthogonal to the surface as well as for a birefringent material with optical axis orientation within the incident plane of the scattering process and under 45° with respect to the sample surface in two different possible cases (see Figs. S12B,C). However, for the results in the main manuscript, the same modelling approach has been followed semi-analytically via a python script for all possible LAO twin-domain walls including canted optical axis orientation.

All **three polarizable spheres** (tip, left and right sample domain) in this model are assumed to be small compared to the wavelength, resulting in induced point-dipoles, which may be represented by their corresponding polarizability tensors. Assuming a metal-coated tip in the mid-IR spectral regime surrounded by air (radius $a$, $\varepsilon_{tip} \ll -1000$ and $\varepsilon_m = 1$), the tip polarizability is given by [13]:

$$\alpha_{tip} = 4\pi a^3 \underbrace{\frac{\varepsilon_{tip} - \varepsilon_m}{\varepsilon_{tip} + 2\varepsilon_m}}_{\approx 1} \approx 4\pi a^3$$

The two sample dipoles on the other hand are represented by uniaxial polarizability tensors $\hat{\alpha}_{LAO}$:

$$\hat{\alpha}_{LAO} = 4\pi b^3 \begin{pmatrix} \eta_\perp & 0 & 0 \\ 0 & \eta_\perp & 0 \\ 0 & 0 & \eta_\| \end{pmatrix}$$

with *b* the sphere radius and

$$\eta_\perp = \frac{\varepsilon_\perp - 1}{\varepsilon_\perp + 2}, \qquad \eta_\| = \frac{\varepsilon_\| - 1}{\varepsilon_\| + 2}$$

including the permittivity component orthogonal ($\varepsilon_\perp$) and parallel ($\varepsilon_\|$) to LAO's optical axis, respectively.

For representation of the twin-domain wall in case A (compare Fig. S12B), the polarizabilities on the two sides of the junction are tilted symmetrically by the angles $\beta_1 = \beta$ and $\beta_2 = -\beta$ (in the calculated example assuming $\beta = 45°$) yielding

$$\hat{\alpha}_1 = \hat{R}(\beta_1) \cdot \hat{\alpha}_{LAO} \cdot \hat{R}^T(\beta_1) = 4\pi b^3 \begin{pmatrix} \frac{1}{2}\cdot(\eta_\perp + \eta_\|) & 0 & \frac{1}{2}\cdot(\eta_\perp - \eta_\|) \\ 0 & \eta_\perp & 0 \\ \frac{1}{2}\cdot(\eta_\perp - \eta_\|) & 0 & \frac{1}{2}\cdot(\eta_\perp + \eta_\|) \end{pmatrix}$$

$$\hat{\alpha}_2 = \hat{R}(\beta_2) \cdot \hat{\alpha}_{LAO} \cdot \hat{R}^T(\beta_2) = 4\pi b^3 \begin{pmatrix} \frac{1}{2}\cdot(\eta_\perp + \eta_\|) & 0 & \frac{1}{2}\cdot(-\eta_\perp + \eta_\|) \\ 0 & \eta_\perp & 0 \\ \frac{1}{2}\cdot(-\eta_\perp + \eta_\|) & 0 & \frac{1}{2}\cdot(\eta_\perp + \eta_\|) \end{pmatrix}$$

with $\hat{R}(\beta)$ being the rotation matrix

$$\widehat{R}(\beta)=\begin{pmatrix} \cos(\beta) & 0 & -\sin(\beta) \\ 0 & 1 & 0 \\ +\sin(\beta) & 0 & \cos(\beta) \end{pmatrix}$$

and $\hat{R}^T(\beta)$ its transposed matrix.

Following the derivation in [8], the **near-field coupling** is implemented by the following steps:

(1) the external field of the incident light induces an initial point dipole in the tip,

(2) the field of the tip dipole at the position of the sample excites virtual sample dipoles

(3) the fields of these virtual sample dipoles modify the field at the position of the tip. Note that these virtual dipoles are assumed not to interact with each other, but solely modify the field at the position of the tip dipole,

(4) the tip dipole itself is modified by the sample-dipole fields resulting in a retro-action that represents the near-field coupling of tip and sample, and

(5) the near-field coupled system is described by a total polarizability tensor $\hat{\alpha}_{tot}$ that includes both modified tip polarizability and sample polarizabilities.

Here, we discuss two different cases of TW, namely cases A and B, as illustrated in Figs. S12B,C. These configurations roughly resemble the 45°-domain walls for (001)-cut sample surfaces which are corresponding to the (110) and $(1\bar{1}0)$ planes of the crystal. Moreover, for simplicity, we assume the positions of the sample spheres to be located at *z*=−*h* and *x* = ±*h*. However, more complex walls and varying angles/sample dipole positions may be addressed with the same model concept as e.g. done via a python script for simulating the responses of all possible LAO domain walls as shown in Fig. 3 of the main manuscript.

These steps towards near-field coupling for the two cases A and B here yield the following:

(1) The initially induced tip dipole is given by

$$\vec{P}_{tip}=\varepsilon_0\cdot\alpha_{tip}\cdot\vec{E}_0\approx\varepsilon_0\cdot 4\pi a^3\cdot E_0\begin{pmatrix}0\\0\\1\end{pmatrix}\qquad \text{at}\qquad \vec{r}_0=\begin{pmatrix}0\\0\\h\end{pmatrix}$$

for an external excitation field $\vec{E}_0$ orthogonal to the sample surface (*z*-direction), and a tip-sample separation of *h=0,785a* used in the python code following values from literature [11, 14].

(2) Its field at the positions of the two polarizable sample spheres may be calculated using

$$\vec{E}(\vec{P},\vec{r})=\frac{1}{4\pi\varepsilon_0}\frac{3\vec{n}\cdot(\vec{P}\cdot\vec{n})-\vec{P}}{|\vec{r}-\vec{r'}|^3}$$

resulting in

$$\vec{E}_{1,tip}=\vec{E}(\vec{P}_{tip},\vec{r}_1)=\frac{a^3}{5^{5/2}h^3}E_0\begin{pmatrix}6\\0\\7\end{pmatrix}\qquad \text{at}\qquad \vec{r}_1=\begin{pmatrix}-h\\0\\-h\end{pmatrix}$$

$$\vec{E}_{2,tip}=\vec{E}(\vec{P}_{tip},\vec{r}_2)=\frac{a^3}{5^{5/2}h^3}E_0\begin{pmatrix}-6\\0\\7\end{pmatrix}\qquad \text{at}\qquad \vec{r}_2=\begin{pmatrix}h\\0\\-h\end{pmatrix}$$

For the two different cases, we assume the following sample configurations (compare

Fig. S12B,C):

$$\underline{\text{Case A}}: \quad \text{Sphere 1:} \quad \vec{r}_1 = \begin{pmatrix} -h \\ 0 \\ -h \end{pmatrix}, \quad \widehat{\alpha}_1$$

$$\text{Sphere 2:} \quad \vec{r}_2 = \begin{pmatrix} +h \\ 0 \\ -h \end{pmatrix}, \quad \widehat{\alpha}_2$$

$$\underline{\text{Case B}}: \quad \text{Sphere 3:} \quad \vec{r}_3 = \vec{r}_1 \qquad \widehat{\alpha}_3 = \widehat{\alpha}_2$$

$$\text{Sphere 4:} \quad \vec{r}_4 = \vec{r}_2, \qquad \widehat{\alpha}_4 = \widehat{\alpha}_1$$

The induced virtual dipoles in the anisotropic spheres 1 to 4 are calculated via the formula

$$\vec{P}_i = \varepsilon_0 \cdot \widehat{\alpha}_i \cdot \vec{E}(\vec{P}_{tip}, \vec{r}_i) \quad \text{with} \quad i = 1, 2, 3, 4$$

with $\vec{E}(\vec{P}_{tip}, \vec{r}_i)$ taken from the above equations for the tip-dipole field at the positions of the sample spheres resulting in the following sample-dipole combinations for the two cases

$$\underline{\text{Case A}}: \quad \text{Sphere 1:} \quad \vec{P}_1 = \varepsilon_0 \cdot 4\pi a^3 \cdot \tfrac{b^3}{2 \cdot 5^{5/2} \cdot h^3} E_0 \begin{pmatrix} 13\eta_\perp - \eta_{||} \\ 0 \\ 13\eta_\perp + \eta_{||} \end{pmatrix} \quad \text{at} \quad \vec{r}_1$$

$$\text{Sphere 2:} \quad \vec{P}_2 = \varepsilon_0 \cdot 4\pi a^3 \cdot \tfrac{b^3}{2 \cdot 5^{5/2} \cdot h^3} E_0 \begin{pmatrix} -13\eta_\perp + \eta_{||} \\ 0 \\ 13\eta_\perp + \eta_{||} \end{pmatrix} \quad \text{at} \quad \vec{r}_2$$

$$\underline{\text{Case B}}: \quad \text{Sphere 3:} \quad \vec{P}_3 = \varepsilon_0 \cdot 4\pi a^3 \cdot \tfrac{b^3}{2 \cdot 5^{5/2} \cdot h^3} E_0 \begin{pmatrix} -\eta_\perp + 13\eta_{||} \\ 0 \\ \eta_\perp + 13\eta_{||} \end{pmatrix} \quad \text{at} \quad \vec{r}_3$$

$$\text{Sphere 4:} \quad \vec{P}_4 = \varepsilon_0 \cdot 4\pi a^3 \cdot \tfrac{b^3}{2 \cdot 5^{5/2} \cdot h^3} E_0 \begin{pmatrix} \eta_\perp - 13\eta_{||} \\ 0 \\ \eta_\perp + 13\eta_{||} \end{pmatrix} \quad \text{at} \quad \vec{r}_4$$

(3) The superposition of the fields of these dipoles at the position of the tip is then given by

$$\underline{\text{Case A}}: \quad \vec{E}_A = \vec{E}(\vec{P}_1, \vec{r}_0) + \vec{E}(\vec{P}_2, \vec{r}_0)$$

$$= \underbrace{\frac{4b^3}{5^5 h^3} \cdot (169\eta_\perp + \eta_{||})}_{=\beta_A''} \cdot \tfrac{1}{16\pi h^3} \cdot \underbrace{\varepsilon_0 \cdot 4\pi a^3 \cdot E_0 \begin{pmatrix} 0 \\ 0 \\ 1 \end{pmatrix}}_{=\vec{P}_{tip}} = \beta_A'' \cdot \tfrac{\vec{P}_{tip}}{16\pi h^3}$$

$$\underline{\text{Case B}}: \quad \vec{E}_B = \vec{E}(\vec{P}_3, \vec{r}_0) + \vec{E}(\vec{P}_4, \vec{r}_0)$$

$$= \underbrace{\frac{4b^3}{5^5 h^3} \cdot (\eta_\perp + 169\eta_{||})}_{=\beta_B''} \cdot \tfrac{1}{16\pi h^3} \cdot \underbrace{\varepsilon_0 \cdot 4\pi a^3 \cdot E_0 \begin{pmatrix} 0 \\ 0 \\ 1 \end{pmatrix}}_{=\vec{P}_{tip}} = \beta_B'' \cdot \tfrac{\vec{P}_{tip}}{16\pi h^3}$$

Note that due to symmetry at the twin-domain wall (with the tip dipole being placed laterally exactly at the position of the wall at *x=y=0*), the sum of the two sample dipole fields at the position of the tip is orthogonal to the surface and with that parallel to the initial tip dipole. Moreover, the factors $\beta_A''$ and $\beta_B''$ are utilized later to calculate the total polarizability.

(4) The induced tip dipole is modified by the fields $\vec{E}_{A/B}$ of the sample dipoles resulting in the near-field coupling via

$$\begin{aligned}\vec{P}_{tip} &= \varepsilon_0 \alpha_{tip} \cdot \left(\vec{E}_0 + \vec{E}_{A/B}\right) \\ &= \varepsilon_0 \alpha_{tip} \cdot \left(\vec{E}_0 + \beta''_{A/B} \cdot \frac{\vec{P}_{tip}}{16\pi h^3}\right)\end{aligned}$$

Solving this equation for $\vec{P}_{tip}$ yields

$$\vec{P}_{tip} = \varepsilon_0 \cdot \frac{\alpha_{tip}}{1 - \frac{\alpha_{tip}\beta''_{A/B}}{16\pi h^3}} \cdot \vec{E}_0$$

(5) The total dipole of the near-field-coupled tip-sample system is represented by the sum of all dipoles, which are assumed to be located at the same position when observed from the far-field ($r \gg h$). The sample dipoles for the two cases A and B therefor can be summarized by

$$\underline{\text{Case A}}: \quad \vec{P}_A = \vec{P}_1 + \vec{P}_2 = \underbrace{\frac{b^3}{5^{5/2} \cdot h^3} \cdot (13\eta_\perp + \eta_\parallel)}_{=\beta'_A} \cdot \underbrace{\varepsilon_0 \cdot 4\pi a^3 \cdot E_0 \begin{pmatrix} 0 \\ 0 \\ 1 \end{pmatrix}}_{=\vec{P}_{tip}} = \beta'_A \cdot \vec{P}_{tip}$$

$$\underline{\text{Case B}}: \quad \vec{P}_B = \vec{P}_3 + \vec{P}_4 = \underbrace{\frac{b^3}{5^{5/2} \cdot h^3} \cdot (\eta_\perp + 13\eta_\parallel)}_{=\beta'_B} \cdot \underbrace{\varepsilon_0 \cdot 4\pi a^3 \cdot E_0 \begin{pmatrix} 0 \\ 0 \\ 1 \end{pmatrix}}_{=\vec{P}_{tip}} = \beta'_B \cdot \vec{P}_{tip}$$

Note that due to the symmetry of the system, the total dipole in both cases is oriented orthogonal to the sample surface, i.e. in $\vec{P}_{tip}$ direction. The resulting total dipole of the near-field coupled tip-sample system then yields

$$\begin{aligned}\vec{P}_{tot} &= \vec{P}_{tip} + \vec{P}_{A/B} \\ &= (1 + \beta'_{A/B}) \cdot \vec{P}_{tip} \quad = \varepsilon_0 \underbrace{\frac{\alpha_{tip}(1 + \beta'_{A/B})}{1 - \frac{\alpha_{tip}\beta''_{A/B}}{16\pi h^3}}}_{=\alpha_{tot}} \vec{E}_0 \quad = \varepsilon_0 \alpha_{tot} \vec{E}_0\end{aligned}$$

In **summary**, the near-field-coupled tip-sample system, hence, is represented by its total polarizability tensor $\hat{\alpha}_{tot}$

$$\alpha_{tot} = \frac{\alpha_{tip}(1 + \beta'_{A/B})}{1 - \frac{\alpha_{tip} \cdot \beta''_{A/B}}{16\pi h^3}}$$

with $\alpha_{tip} = 4\pi a^3$ being the polarizability of the spherical metal tip of radius *a*, *h* being the separation between tip dipole and sample surface (typically set to *h=0.785a* [11, 14]) and the complex functions $\beta'_{A/B}$ as well as $\beta''_{A/B}$ being different for the two twin-domain wall types (Cases A and B as shown in Fig. S12B,C) and given by

$$\underline{\text{Case A}}: \qquad\qquad \underline{\text{Case B}}:$$

$$\beta'_A = \frac{b^3}{5^{5/2}h^3} \cdot (13\,\eta_\perp + \eta_\parallel) \qquad \beta'_B = \frac{b^3}{5^{5/2}h^3} \cdot (\eta_\perp + 13\,\eta_\parallel)$$

$$\beta''_A = \frac{4b^3}{5^5 h^3} \cdot (169\,\eta_\perp + \eta_\parallel) \qquad \beta''_B = \frac{4b^3}{5^5 h^3} \cdot (\eta_\perp + 169\,\eta_\parallel)$$

Both $\beta'_{A/B}$ and $\beta''_{A/B}$ are functions of $\eta_\perp$ and $\eta_\parallel$, which depend on the complex optical sample permittivity components orthogonal ($\varepsilon_\perp$) and parallel ($\varepsilon_\parallel$) to its optical axis via

$$\eta_\perp = \frac{\varepsilon_\perp - 1}{\varepsilon_\perp + 2}, \qquad \eta_\parallel = \frac{\varepsilon_\parallel - 1}{\varepsilon_\parallel + 2}$$

The total polarizability tensor may be used to **<u>calculate the different signals measured by s-SNOM</u>** such as the complex scattered electric field $\vec{E}_{sca}$, scattering ($C_{sca}$) and absorption ($C_{abs}$) cross sections. The corresponding formulae are given by:

$$\begin{aligned} \vec{E}_{sca} &\sim \alpha_{tot} \cdot \vec{E}_0 \\ C_{sca} &= \frac{k^4}{6\pi} \cdot |\alpha_{tot}|^2 \\ C_{abs} &= k \cdot \Im m(\alpha_{tot}) \end{aligned}$$

Typically, in order to determine the resonances of a near-field coupled tip-sample system, $C_{sca}$ is discussed as function of wavenumber. Figure S13 shows $C_{sca}$ as function of wavenumber for the two different cases of twin domain walls in LAO. Enhanced signals in s-SNOM are typically observed around sample permittivities of $\varepsilon \approx -5 \ldots -1$ with the exact spectral location being dependent on the tip-sample separation $h$. The corresponding enhancement is observed in all three Reststrahlen bands matching the experimental observation and numerical results. Due to the tilted optical axis within the domains (see Fig. S12), the two different twin-domain wall configurations (cases A and B) show resonances that are dominated either by the perpendicular or the parallel component of the dielectric tensor, depending on the axis alignment with respect to the tip-dipoles electric field at the location of the sample dipoles. This effect is illustrated in Fig. S14 by sketching the electric field lines of the tip dipole at the positions of the sample spheres.

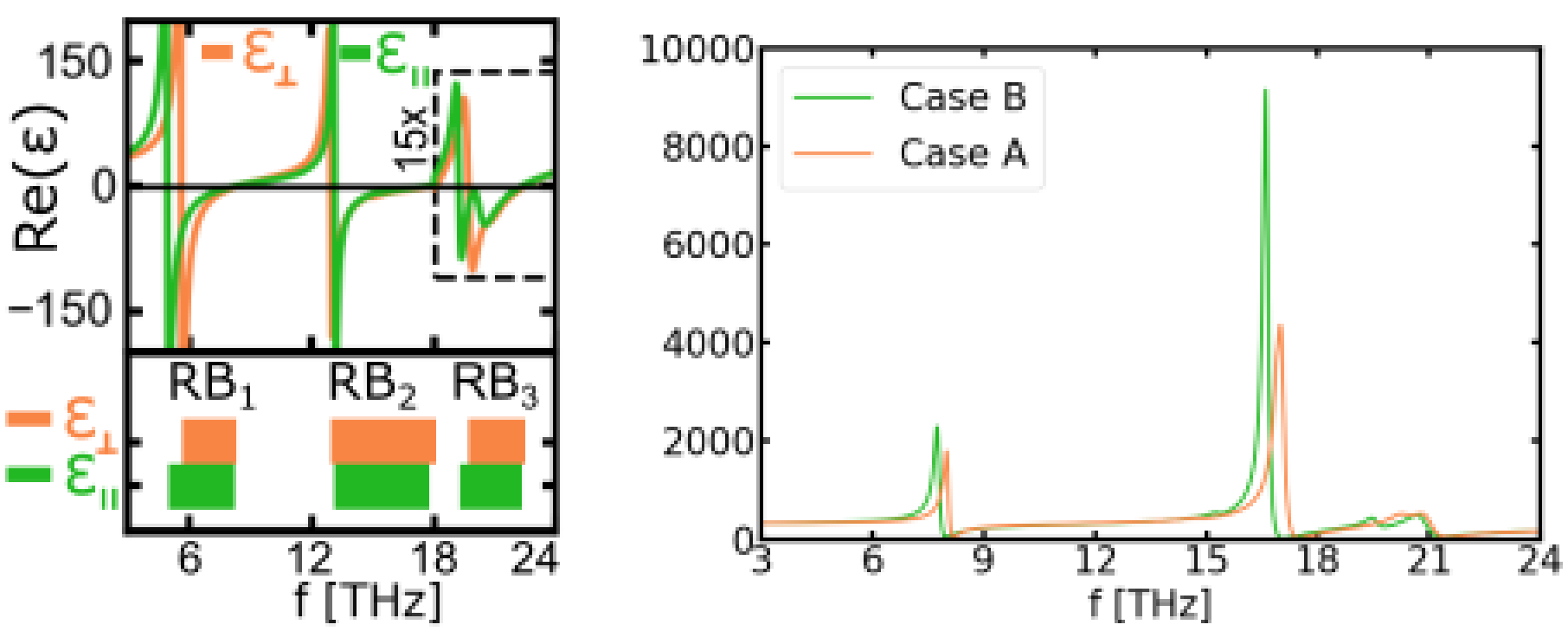

**Figure S13.** Permittivity data (left) and corresponding scattering cross sections (right) calculated via the here developed dipole model for twin domain walls for the two different cases as displayed in fig. S12.

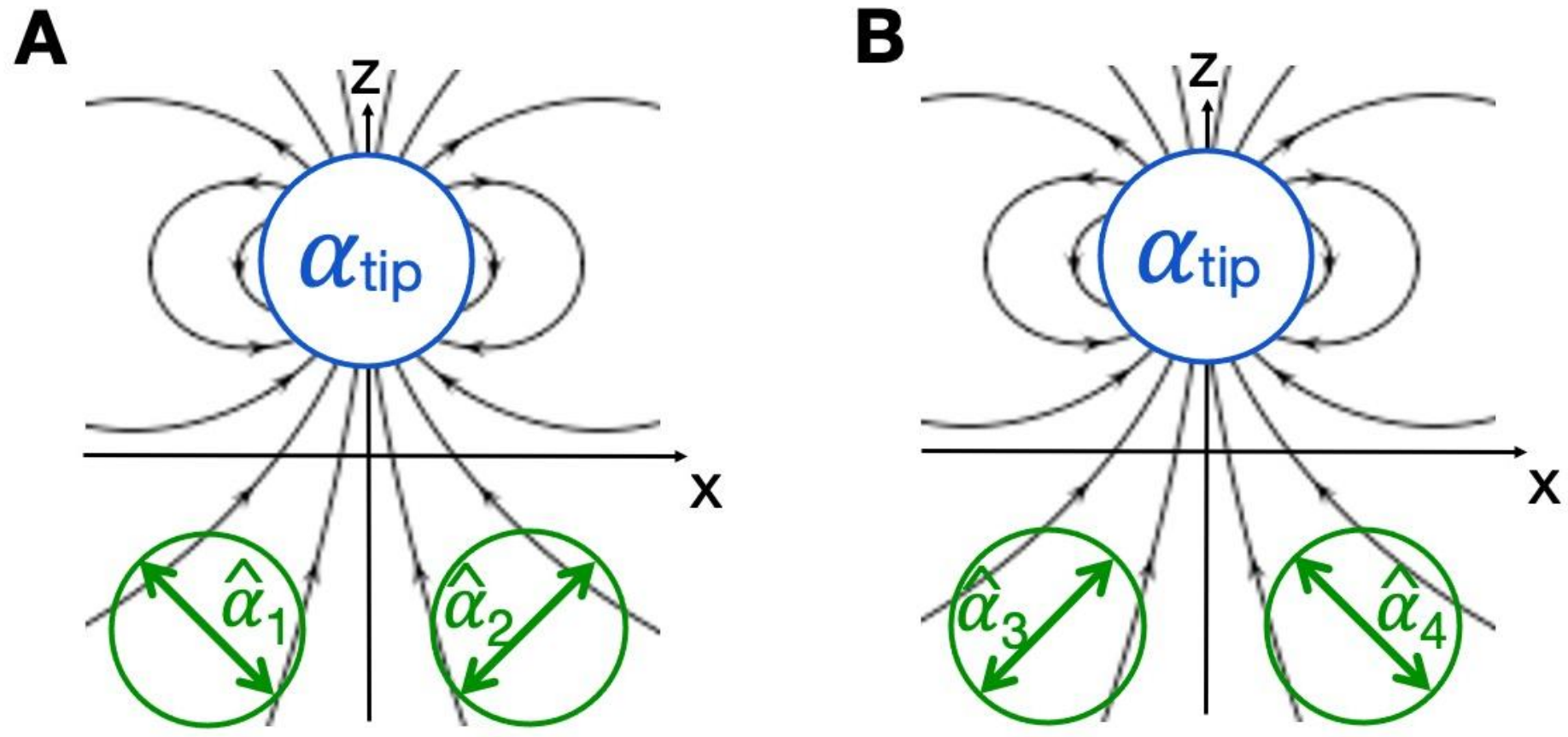

**Figure S14.** Illustration why the model predicts probing of the two different permittivity components of the sample for the two different cases. **A** In case A, the field lines of the tip dipole at the position of the sample dipoles are oriented mostly orthogonal to the optical axis in both domains, whereas **B** in case B, the field lines are oriented parallel to the optical axis.

In conclusion, we modified the analytical point dipole model for probing of domain walls between two sample areas with different permittivity tensors. Therefore, the sample responses of the two areas next to the tip, which is located exactly above the domain wall, are represented by anisotropic, polarizable spheres. The permittivity tensor of these spheres in principle can be chosen at will in the developed model, however here we focus on a symmetric configuration as being present for twin domain walls e.g. in (001)-cut LAO. Matching approximately the 45° twin-domain walls in (001)-cut LAO, we discuss two different cases of twin domain walls with different orientations of the optical axis and find resonances that are dominated by the permittivity component orthogonal and parallel to the optical axis for the two cases A and B, respectively. This effect is determined mathematically by the total polarizability of the near-field coupled system, which shows a resonant maximum when its denominator approaches zero.

This simple analytical model enables us to qualitatively describe the near-field signal and domain-wall contrast found experimentally on LAO within its Reststrahlen bands matching perfectly the rastering-dipole COMSOL simulations shown in Figure 2 of the Main. Whereas here, we limit ourself quite strongly in terms of polarisation of the excitation field and optical axis orientation in the domains, the model in general holds for arbitrary choices of both, which can be easily implemented e.g. in a python script which enables the calculation of the coupling without restrictions on these parameters.